\documentclass[%
amsmath,amssymb,aps,prb,twocolumn,
superscriptaddress]{revtex4-2}

\usepackage{graphicx}% Include figure files
\usepackage{dcolumn}% Align table columns on decimal point
\usepackage{bm}% bold math
\usepackage{xcolor}
\usepackage{verbatim}
\usepackage{epsfig}

\begin{document}

\preprint{APS/123-QED}

\title{Strong effects of thermally induced low-spin-to-high-spin crossover on transport properties of correlated metals}

\author{Johanna Moser}
\affiliation{Institute of Theoretical and Computational Physics, TU Graz, NAWI Graz, Petersgasse 16, 8010 Graz, Austria}
\author{Jernej Mravlje}%
\affiliation{Jožef Stefan Institute, Jamova 39, SI-1000 Ljubljana, Slovenia}
\affiliation{Faculty of Mathematics and Physics, University of Ljubljana, Jadranska 19, 1000 Ljubljana, Slovenia} 

\author{Markus Aichhorn}
\email[Corresponding author: ]{aichhorn@tugraz.at}
\affiliation{Institute of Theoretical and Computational Physics, TU Graz, NAWI Graz, Petersgasse 16, 8010 Graz, Austria}

\date{\today}% It is always \today, today,

\begin{abstract}
  We use dynamical mean-field theory to study how electronic transport
  in multi-orbital metals is influenced by correlated (nominally) empty orbitals
  that are in proximity to the Fermi level. Specifically, we study $2+1$ orbital and $3+2$ orbital (i.e. $t_{2g} +e_g$) models on  a Bethe lattice with a crystal field that is set so that the higher lying orbitals are nearly empty at low temperatures but get a non-negligible occupancy at elevated temperature. The high temperature regime is characterized by thermal activation of carriers leading to higher magnetic response (i.e., thermally induced low-spin to high-spin transition) and substantial influence on resistivity, where one can distinguish two counteracting effects: increased scattering due to formation of high spin and increased scattering phase space on one hand, and additional parallel conduction channel on the other. The former effect is stronger and one may identify cases where resistivity increases by a factor of three at high temperatures even though the occupancy of the unoccupied band remains small ($<10\%$). We discuss implications of our findings for transport properties of correlated materials.
\end{abstract}

\maketitle

%\tableofcontents

\section{Introduction}
How resistivity depends qualitatively on temperature provides a basic characterization of electronic state in solids: a resistivity $\rho$ that drops with temperature indicates a semiconducting state,  a $\rho \propto T^2$ behavior indicates electron-electron scattering dominated metallic Fermi-liquid behavior~\cite{landau,imada1998}, certain power laws indicate proximity to distinct magnetic instabilities~\cite{lohneysen2007}, the $T-$linear resistivity indicates electron-phonon dominated~\cite{ziman} simple metallic conduction (but also electron-electron dominated strange-metal behavior notorious in cuprates). More rarely (but attracting increasingly more attention) the quantitative values of resistivity are used to characterize the behavior as well. One such example is the Mott-Ioffe-Regel minimal conduction criterion distinguishing good and bad metals~\cite{gunnarsson2003,deng2013}. Recently significant attention was attributed to the prefactor of $T-$linear resistivity and argued to be often such that the scattering rate is ``Planckian''~\cite{bruin2013,michon2023,georges2021}

Clearly, in calculations of transport within specific models the obtained magnitude of scattering should be given the necessary attention. For instance, SrVO$_3$ that was always considered to be in an electron-electron scattering dominated Fermi liquid state was recently shown to be actually dominantly affected by phonon scattering~\cite{abramovitch2024respective}. 
On the applied side likewise the magnitude of scattering is important, for instance in applications of transport calculations in geophysics the electron-electron vs. electron-phonon scattering are widely debated~\cite{xu2018,pourovskii2020,zhang2021,blesio2023influence}. 

Even if one considers the electron-electron interactions alone, the details of the considered model may have a significant influence on the result.
Namely, in 
transition metal oxides where the $t_{2g}$ manifold is partially occupied, one often disregards $e_{g}$ states (see Ref.~\cite{abramovitch2023} for a recent example) for simplicity and computational efficiency. But this may become increasingly inaccurate at elevated temperatures because the bottom of the $e_g$ band is not far above the Fermi level and is eventually populated by thermally activated electrons, which are strongly coupled to $t_{2g}$ electrons through the Hund's rule coupling. One can envisage different counteracting influences of these nearly empty bands (which we will call minority bands from now on): (i) the electrons in those minority bands can increase the scattering (either due to additional phase space for the final states, or via additional interactions, or due to the changes in occupancy of the majority band manifold, bringing it closer to half filling).  (ii) electrons in minority bands might scatter less due to the small electron density and hence act as efficient parallel conduction channel leading to increased conduction. 

Due to Hund's coupling, the thermal population of $e_g$ states will lead to an increased magnetic response. This phenomenon is known as spin-crossover or low-spin-to-high-spin transition~\cite{book2004,lowspin_highspin_propositions1} and has been widely explored in  transition-metal organic complexes~\cite{Gtlich2000,SCO_MOFs, SCO_conductivity_2023, ewald_crossover, SCO_CP_conductor}, in d$^6$ oxides such as LaCoO$_3$~\cite{abbate93,korotin96,eder2010,krapek2012,chakrabarti2017,takegami2023} and in iron periclase~\cite{Wentzcovitch2010,tsuchiya06} materials, to name a few examples. In the quoted cases the phenomenon is strong and involves large redistribution of the carriers and is accompanied by large structural changes. 

Of interest to the present work is the onset of phenomena when the occupation of $e_g$'s remain mild, say less than 10\%. Might the onset of formation of large spin importantly affect the scattering, the resistivity, and magnetic response even there? 
A specific example where all these effects might play a role are early transition metal oxides, such as ruthenates~\cite{mackenzie03} and rhodates~\cite{Perry2006}. Especially ruthenates were extensively studied in dynamical mean-field theory~\cite{mravlje11,behrmann12,dang15,dang15prb,mravlje16,han16,dasari16,zhang16,kondo16,kim18,facio18,tamai19,strand19,sarvestani19,kugler20,linden20,lee21,kaeser22,Shorikov2022,Suzuki2023,moon23,blesio24} with calculations with exception of~\cite{kotliar_transport_in_ruthenates} being done within the $t_{2g}$ subspace. 

In this paper we investigate these questions in the context of a simple model with $2+1$ and $3+2$ orbitals, with the notation $M+M'$, where $M$ indicates the degeneracy of the majority manifold at the Fermi level, and $M'$ the minority orbital manifold, respectively. The crystal field is tuned such that the higher lying minority orbitals are practically empty at very low temperatures, but get a non-negligible occupancy at elevated temperature. We solve the model with Dynamical Mean-Field Theory (DMFT) and find that especially for strong interactions, an additional coupled orbital does indeed have a strong influence on resistivity, mainly by introducing additional scattering for the majority carriers which has a larger influence than the opening of an additional conduction channel. 
We want to note that earlier DMFT model studies have explored the effects of crystal field, but mainly focusing on the ground state phase diagram~\cite{werner07,kunes_augustinsky_2014,kunes2014,hoshino16}.

\section{Methods}

\begin{figure}[t]
\centering
\includegraphics[width=0.9\columnwidth]{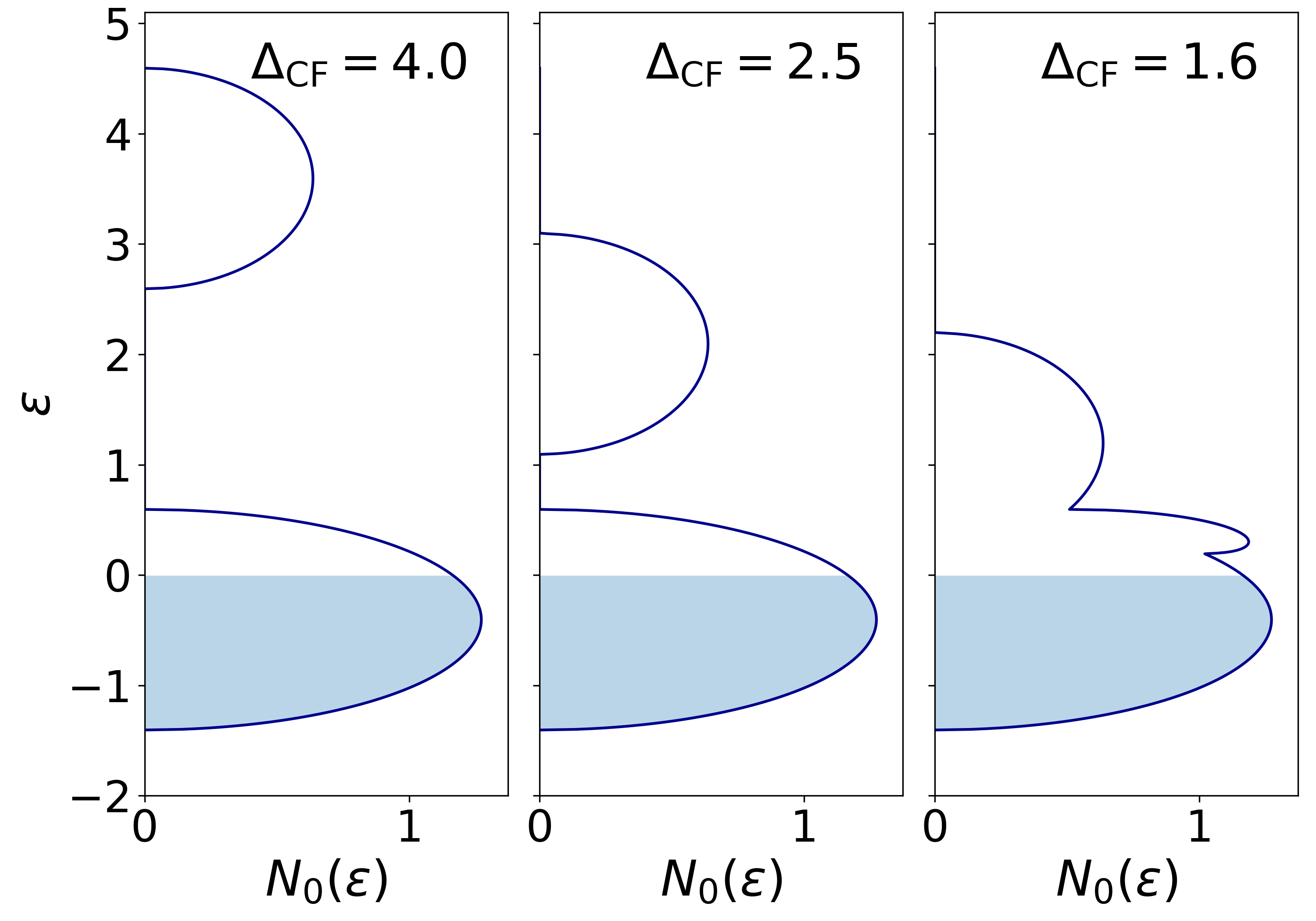}
	\caption{Non-interacting density of states for the 2+1 orbital model on the Bethe lattice for crystal fields $\Delta_\mathrm{CF} = 4$ (left), $\Delta_\mathrm{CF} = 2.5$ (middle), and $\Delta_\mathrm{CF} = 1.6$ (right). The filling at $T = 0$\,K is depicted in light blue color.}
	\label{fig:im_CF_DOS}
\end{figure}

We consider a multi-orbital problem on a Bethe lattice described by a semicircular density of states $\rho_m(\epsilon) = 2/(\pi D) \sqrt{1-(\epsilon/D)^2}$ for each of the orbital subspaces $m$, leading to the kinetic part of the Hamiltonian $H_\mathrm{kin}$. We also consider a term that breaks crystal-field degeneracy; it increases the energy for the orbitals in the high-energy subspace, see also Fig.~\ref{fig:im_CF_DOS}. Hence, the noninteracting  Hamiltonian can be written as    
\begin{equation}\hat{H}_\textrm{0} = \hat{H}_\textrm{kin} + \Delta_\mathrm{CF} \sum_{n \in M', \sigma} \hat{n}_{n\sigma} \, .
\end{equation}
For the interaction term, we use the Kanamori Hamiltonian~\cite{hund_coupling_jernej} parametrized by Coulomb interaction $U$ and Hund's coupling $J$, 
\begin{equation}\label{eq:kanamori}
 \begin{split}
H_\textrm{int}& = U\sum_m \hat{n}_{m\uparrow} \hat{n}_{m\downarrow} + U'\sum_{m\neq m'}  \hat{n}_{m\uparrow} \hat{n}_{m'\downarrow} \\
&\quad+ (U'-J)\sum_{m<m', \sigma}\hat{n}_{m \sigma} \hat{n}_{m' \sigma} \\
&\quad+J\sum_{m\neq m'}c_{m\uparrow}^\dag c_{m'\downarrow}^\dag c_{m\downarrow} c_{m'\uparrow}\\
&\quad + J\sum_{m\neq m'} c_{m\uparrow}^\dag c_{m\downarrow}^\dag c_{m'\downarrow} c_{m'\uparrow}.
 \end{split}
\end{equation}
The indices $m, m'$ run over all $M+M'$ orbitals, and we set $U' = U-2J$ to make the Hamiltonian rotationally invariant in orbital space. 
We solve the impurity model using the CTHYB solver~\cite{cthyb_long,cthyb_triqs} as implemented in the TRIQS package~\cite{triqs}. We performed analytic contination of the Matsubara self energies $\Sigma(i \omega_n)$ to the real frequency axis using the stochastic Maximum Entropy method~\cite{stochME} and Padé continuations.
The considered density of states is schematically presented in Fig.~\ref{fig:im_CF_DOS} for a few values of crystal field. 

The main quantity of interest in this work is the electric conductivity $\sigma$. 
On a Bethe lattice, it is given~\cite{deng2013,arsenault,cox,vertexcorrections, chung} by 
\begin{equation}
\sigma =\frac{2\pi e^2}{\hbar} \sum_{\nu}\int_{-\infty}^{+\infty}d\epsilon \Phi(\epsilon)\int_{-\infty}^{+\infty} d\omega \left( -\frac{\partial f(\omega)}{\partial \omega}\right) A_{\nu, \epsilon}^2(\omega)\, ,
\label{eq:bubble}
\end{equation}
where 
\begin{equation}
\Phi(\epsilon) = \frac{2}{3\pi D^2}\Theta(D - |\epsilon|)(D^2 - \epsilon^2)^{\frac{3}{2}}
\label{eq:transportfunction}
\end{equation}
is the transport function on the Bethe lattice, and
%\added[id=jm]{(Are we sure about $\pi'$, factors of 2, summations over spins)}, and 
\begin{equation}
A_{\nu, \epsilon}(\omega) = -\frac{1}{\pi} \textrm{Im} \left( \omega +i0^+ - \epsilon + \mu - \Sigma_{\nu}(\omega) \right)^{-1}
\end{equation}
is the spectral function of each orbital $\nu$ that depends on band energy $\epsilon$. 
The calculations were done in paramagnetic state: we are interested in the high temperature regime above any possible magnetic ordering temperature. 
The integrals necessary to obtain the temperature dependent resistivity were numerically calculated using an integrator based on globally adaptive interval subdivision \cite{scipy}. In the remainder of the paper we will give all energies in units of the half band width $D$ of the Bethe lattice, and the conductivity in units of $\sigma_0 = e^2 \Phi(0)/(\hbar D)$ \cite{deng2013}.

In order to get insight into the degree of correlation in the different regimes, we extracted the quasiparticle scattering rate through extrapolation of the Matsubara self energies to zero through a polynomial fit, 
\begin{equation}\tau^{-1} = -Z\, \textrm{Im}\,\Sigma(i\omega_n \rightarrow 0)\, ,
\label{eq:scatteringrates}
\end{equation}
where $Z$ is the quasiparticle weight 
\begin{equation}\label{eq:quasiparticle} Z =   \left(1 -{ \frac{\partial}{\partial \omega} \textrm{Im}\Sigma(i\omega)} |_{\omega \rightarrow 0^+} \right) ^{-1}\, .
\end{equation}
Strictly speaking, this definition is valid in the low-temperature limit. However, we adopt it also at higher temperatures, as long as it gives physicaly sensible results, i.e., $0\leq Z\leq 1$. 

We relate the behavior in transport also to the magnetic response that we monitor using the local spin susceptibility 
\begin{align}
\chi_\mathrm{loc} &=\int_0^{\beta} d\tau \langle S^z(\tau) S^z(0)\rangle\, , \\ \label{eq:magsus}
S^z(\tau) &= \frac{1}{2}\sum_{\nu}\left( \hat{n}_{\uparrow \nu}(\tau) - \hat{n}_{\downarrow \nu}(\tau)\right)\, ,
\end{align}	
where $\beta= 1/T$ is the inverse temperature (we set $k_B=1$). The correlation function can be measured directly during the QMC solution of the Anderson impurity problem~\cite{cthyb_triqs}.

\begin{figure}[t]
\centering
\includegraphics[width=0.9\columnwidth]{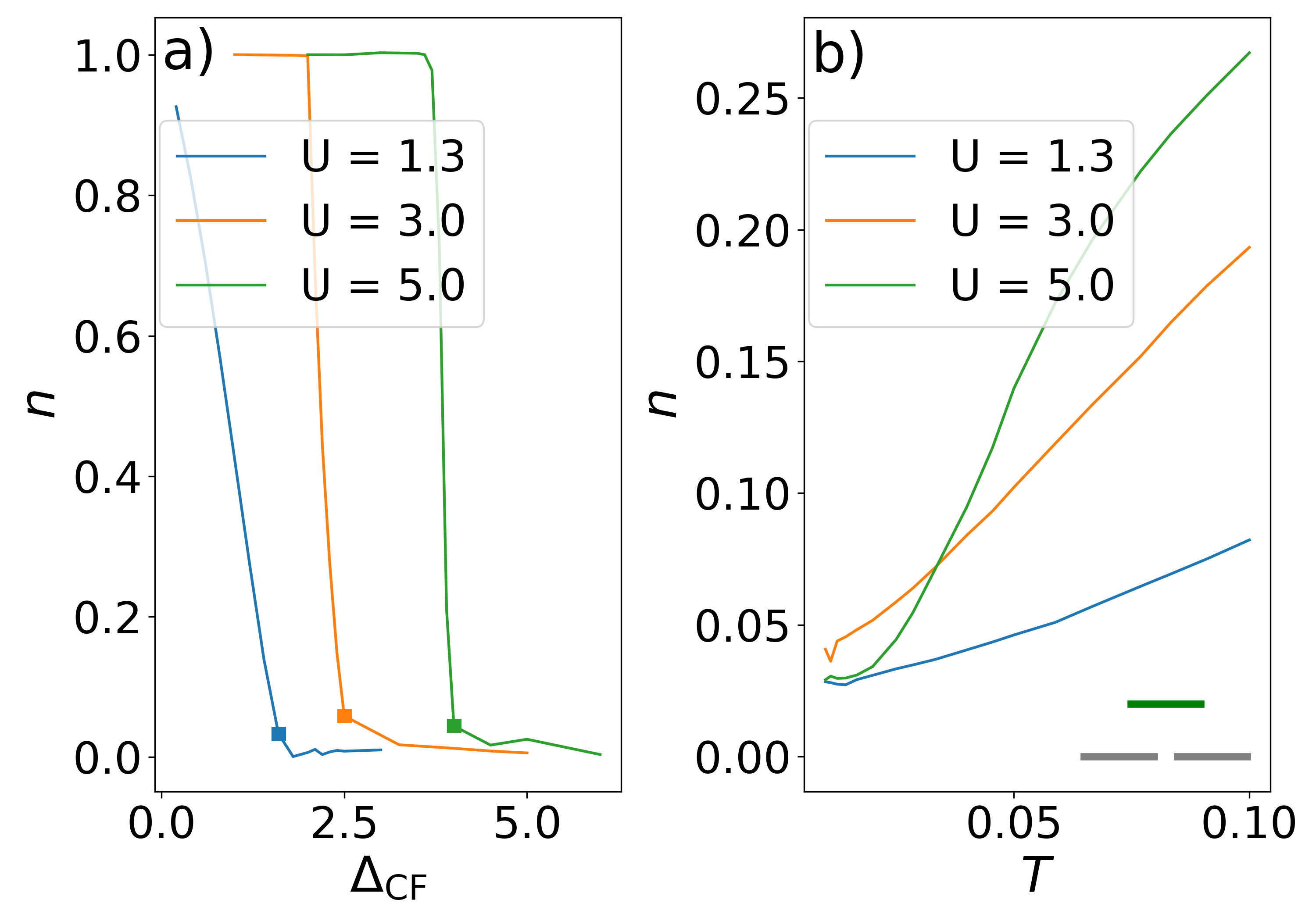}
	\caption{a) Occupancy of the minority orbitals when varying the crystal field at $T=0.025$ ($\beta = 40$). b) Occupancy of the minority orbitals with rising temperatures at fixed crystal field indicated by the squares in a) for respective values of interaction.}
  	\label{fig:im_occupation_TCF}
\end{figure}

\section{2+1 orbital model}
First, we consider a 2+1 orbital model.
We set the total occupancy to a total number of $N=3$ electrons. We consider three different interaction strengths $U= 5.0, 3.0, 1.3$ and fix $J/U = 0.2$. The calculated dependence of occupancy of the minority orbital $n$ on the crystal field $\Delta_\mathrm{CF}$ is shown in Fig.~\ref{fig:im_occupation_TCF}~a). One sees a crossover from the fully orbitally polarized (low-spin) to an unpolarized (high-spin) situation that becomes abrupt for large interactions. In the atomic limit, the criterion for the crossover is given by equating the crystal field energy cost to the Hund's energy gain which for the 2+1 orbital model  ($S=3/2$ vs. $S=1/2$ state) reads $\Delta_\mathrm{CF}=4J$. One sees that this criterion indeed explains the behavior.

Next, respectively for each $U$ value, we choose the crystal fields close to the crossover point $\Delta_\mathrm{CF}=4.0, 2.5, 1.6$ marking the corresponding points by squares in Fig.~\ref{fig:im_occupation_TCF}~a) and plot the electronic density (occupancy) in the minority orbital $n(T)$ in Fig.~\ref{fig:im_occupation_TCF}~b). The occupancy of the nominally empty orbital, which is small at low temperatures, grows significantly with temperature and the growth is more rapid for larger interaction strengths. 

\begin{figure}[t]
\centering
	\includegraphics[width=0.95\columnwidth]{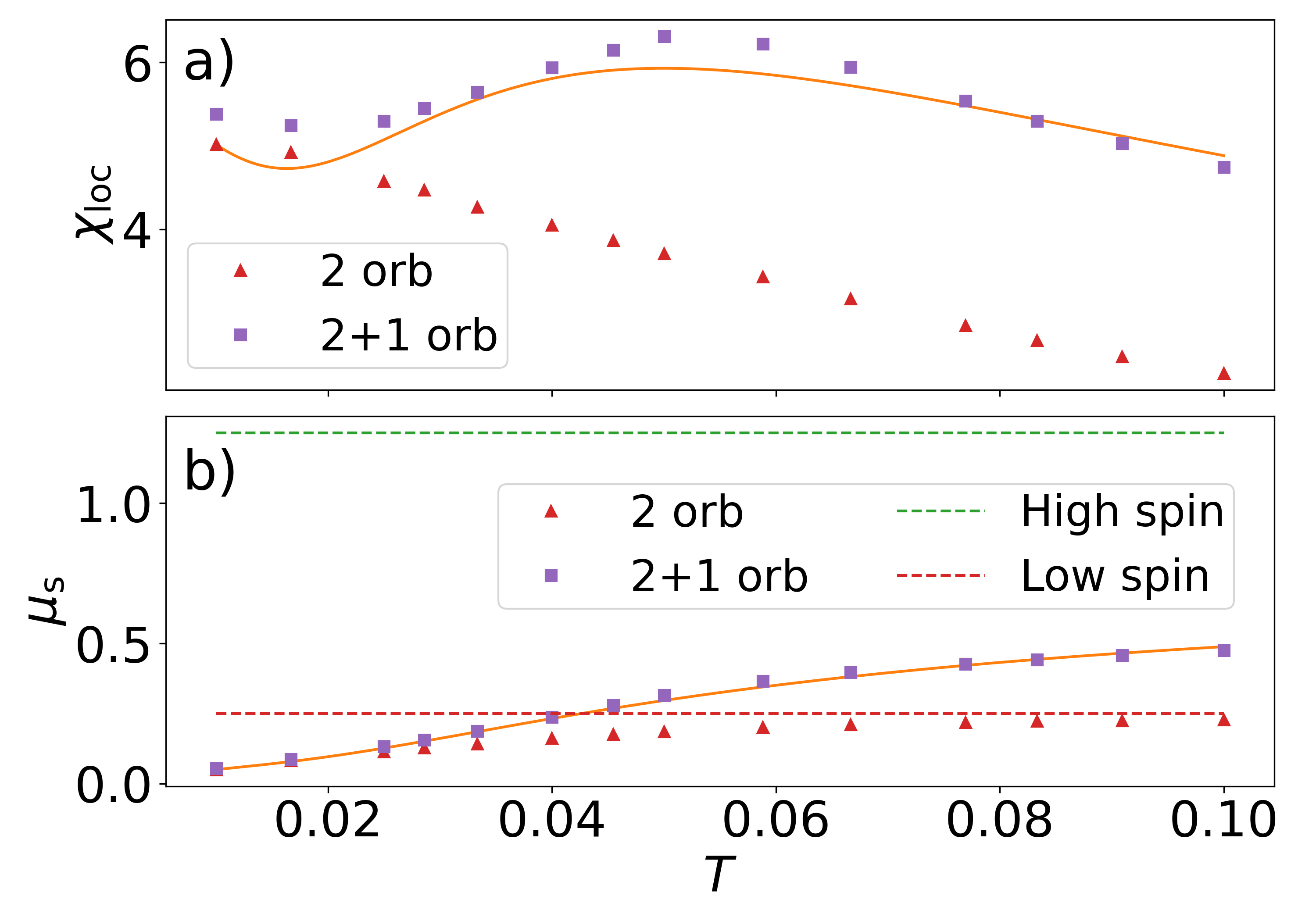}
	\caption{a) Local magnetic susceptibility and magnetic moment as a function of temperature for $U = 5.0, \Delta_\mathrm{CF}=4.0$. In a) we show the local magnetic susceptibility of the 2 orbital model (red triangles), and of the 2+1 orbital model (violet squares) calculated using Eq.~\eqref{eq:magsus}, as well as a corresponding fit after eq. \ref{eq:susfit} that is discussed in detail in App.~\ref{sec:appA}. In panel b), the corresponding (spin dependent) magnetic moment $\mu_\mathrm{s}=\chi_\mathrm{loc} T $ is plotted. The dashed horizontal lines mark the saturated moments, red for the 2 orbital model, green for the 2+1 orbital model.
 }
	\label{fig:im_susceptibility}
\end{figure}

In Fig.~\ref{fig:im_susceptibility}~a), we plot the local susceptibility of the 2 and 2+1 orbital models at $U = 5.0, \Delta_\mathrm{CF}=4.0$ (symbols) as function of temperature.
Whereas the magnetic susceptibility of the 2-orbital model falls monotonously with $T$ and is roughly described by a Curie-Weiss law, the one calculated for the 2+1 orbital model is non-monotonous and has a clear maximum at $\beta \approx 20$. 
One can fit the behavior quite well using a formula with two fit parameters only, eq. \ref{eq:susfit}. A discussion on the fit procedure and the parameters can be found in App.~\ref{sec:appA}. The equation is in agreement with early general results of Van~Vleck~\cite{vanvleck_original}, where a general expression (eq. \ref{eq:VanVleck}) for a susceptibility with contributions from several multiplets split by energies $E_S$ in an atom has been derived. Each of these multiplets show a Curie-Weiss like behavior, but the contribution of energetically higher orbitals is thermally activated, thus giving rise to the possibility of a non-monotonic susceptibility, such as in Fig. \ref{fig:im_susceptibility}.  Materials showing such non-monotonous behavior have been discussed for example in \cite{mulak, vanvleck_koenig, ewald_crossover}, and the appearance of such a susceptibility has even been suggested to be an experimentally accessible way of evaluating if a low-spin high-spin transition occurs \cite{ewald_crossover}. Note, however, that the existence of a local maximum is is not a necessary feature of such a transition, as we will see for example in Fig. \ref{fig:im_suszept_5orb}~a). 
It is interesting that this effect is apparent already for a moderate occupancy ($n\sim 0.15$ at the maximum) of the minority orbital. 
Fig.~\ref{fig:im_susceptibility}~b) shows the paramagnetic moment $\mu_\mathrm{s}=\chi_\mathrm{loc} T $. One sees a clear enhancement of $\mu_\mathrm{s}$ over the $S=1/2$ value of $1/4$ (indicated by the red dashed horizontal line), but the system is still far from the full HS behavior that is characterized by $\mu_\mathrm{s} =5/4$ (indicated by the green dashed line).

\begin{figure}[t]
\centering
\begin{minipage}[t]{0.99\columnwidth}
\centering
	\includegraphics[width=1.0\textwidth]{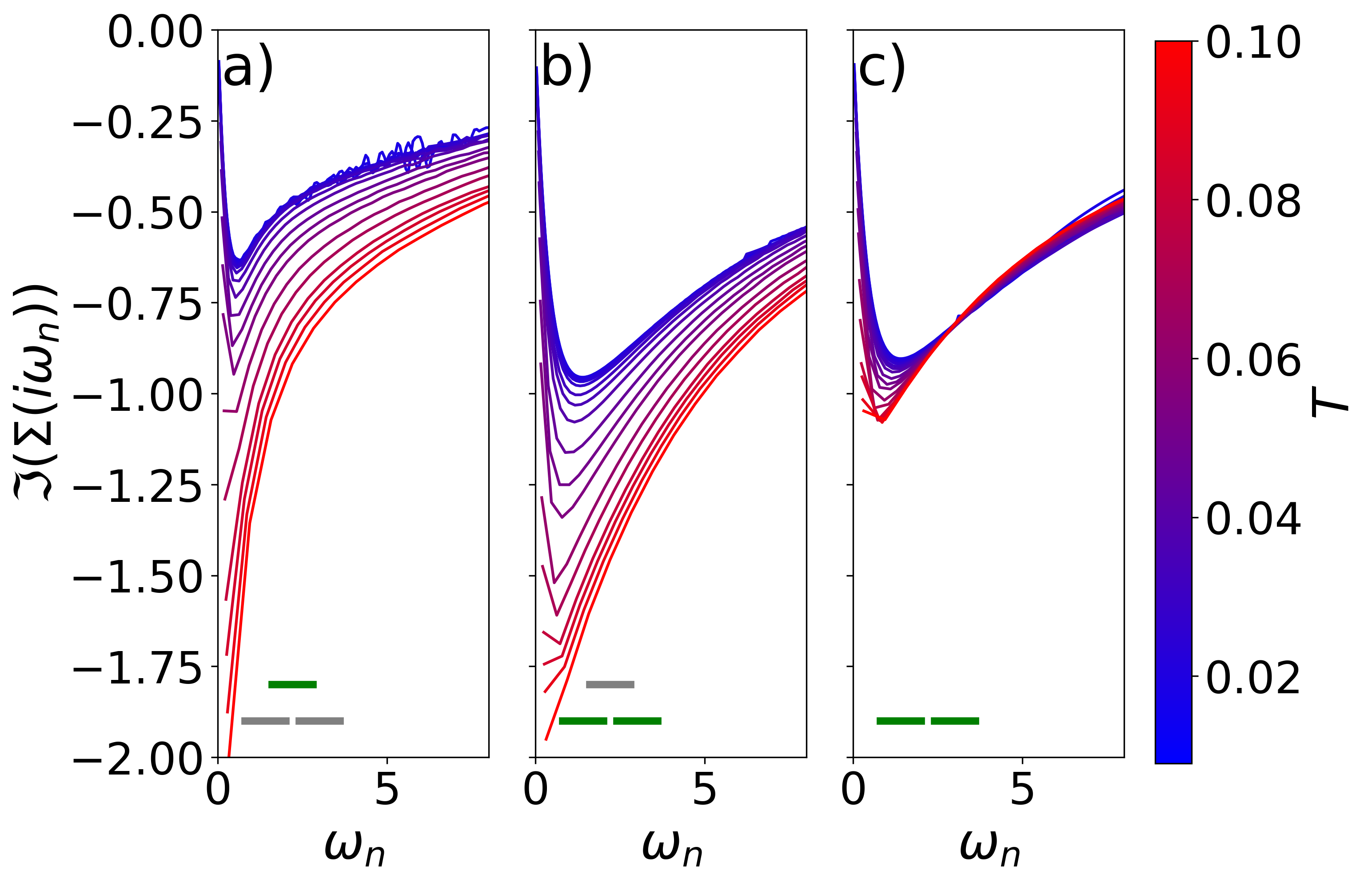}
 \end{minipage}
\begin{minipage}[t]{0.99\columnwidth}
\centering
	\includegraphics[width=1.0\textwidth]{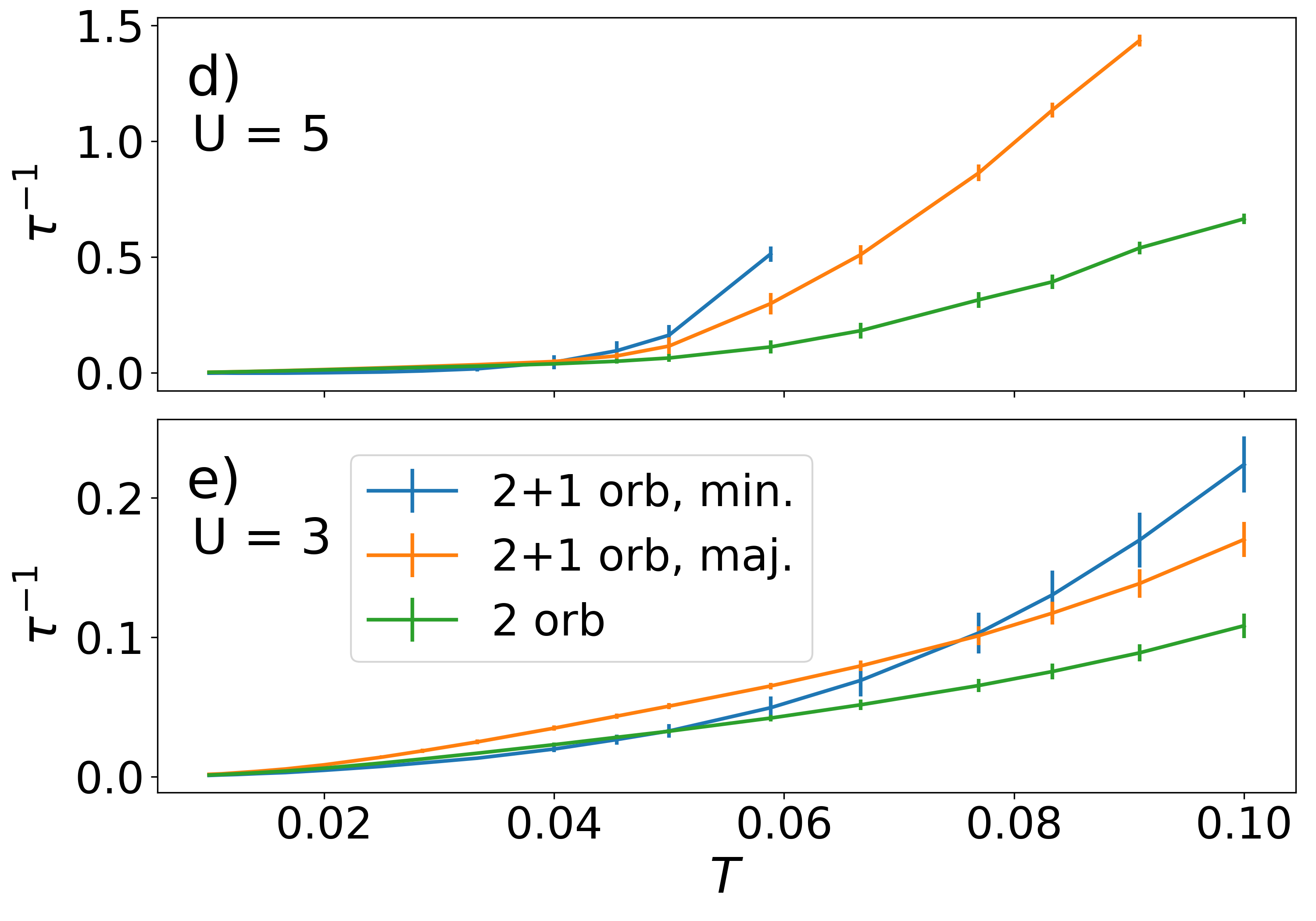} %old version: im_scattering
 \end{minipage}
	\caption{a) The imaginary part of the Matsubara self energies at $U = 5.0$, $\Delta_\textrm{CF}=4.0$ for several $T$ indicated by a color gradient from blue ($T = 0.01$) to red ($T = 0.1$) for a) minority and b) majority orbital of the 2+1 orbital model. c) The two orbital model result.  (d,e) Quasiparticle scattering rate for $U=5.0$, $\Delta_\textrm{CF}=4.0$, and $U=3.0$, $\Delta_\textrm{CF}=2.5$, resp. We are showing data only up to temperatures where the definition of the quasi-particle renormalization is valid, see eq.~\eqref{eq:quasiparticle}.}
	\label{fig:im_selfenergies_paper}
\end{figure}

How does the presence of the additional orbital affect scattering? In Fig.~\ref{fig:im_selfenergies_paper} we show the Matsubara self energies for the 
minority (a) and the majority orbital (b). We compare them to the 2 orbital model result (c). The data is compiled in Fig.~\ref{fig:im_selfenergies_paper}~(d,e) where we plot the temperature dependence of the quasiparticle scattering rate for $U=5.0$, $\Delta_\textrm{CF}=4.0$, and $U=3.0$, $\Delta_\textrm{CF}=2.5$, respectively. 

At low temperatures, the self energies of the majority  orbitals of the 2+1 orbital model match the ones of the 2 orbital model but upon increasing temperature, the magnitude of the self energy in the 2+1 case increases more rapidly. At high $T$ the 2+1 orbital result indicates a significantly more correlated state. This behavior likely results from the more strongly correlated high-spin states, even though the self energies are significantly smaller than the ones found in the unpolarized three-orbital model (i.e., $\Delta_\textrm{CF}=0.0$) at half filling. One can also interpret it in terms of reduced occupancy in the majority orbitals, although the filling $n\approx 2.8$ in these orbitals is still far away from the majority orbital half filling of $n=2.0$.

Interestingly, the scattering in minority orbital is quite high and is larger than what one would think considering the small number of electrons there (the imaginary part $|\mathrm{Im} \Sigma|$ of a single-orbital model in the same low-density regime is an order of magnitude smaller). 
Similar large scattering in nominally unactive orbitals has been seen in nickelates~\cite{deng2012}.

\begin{figure}[t]
\centering
	\includegraphics[width=0.95\columnwidth]{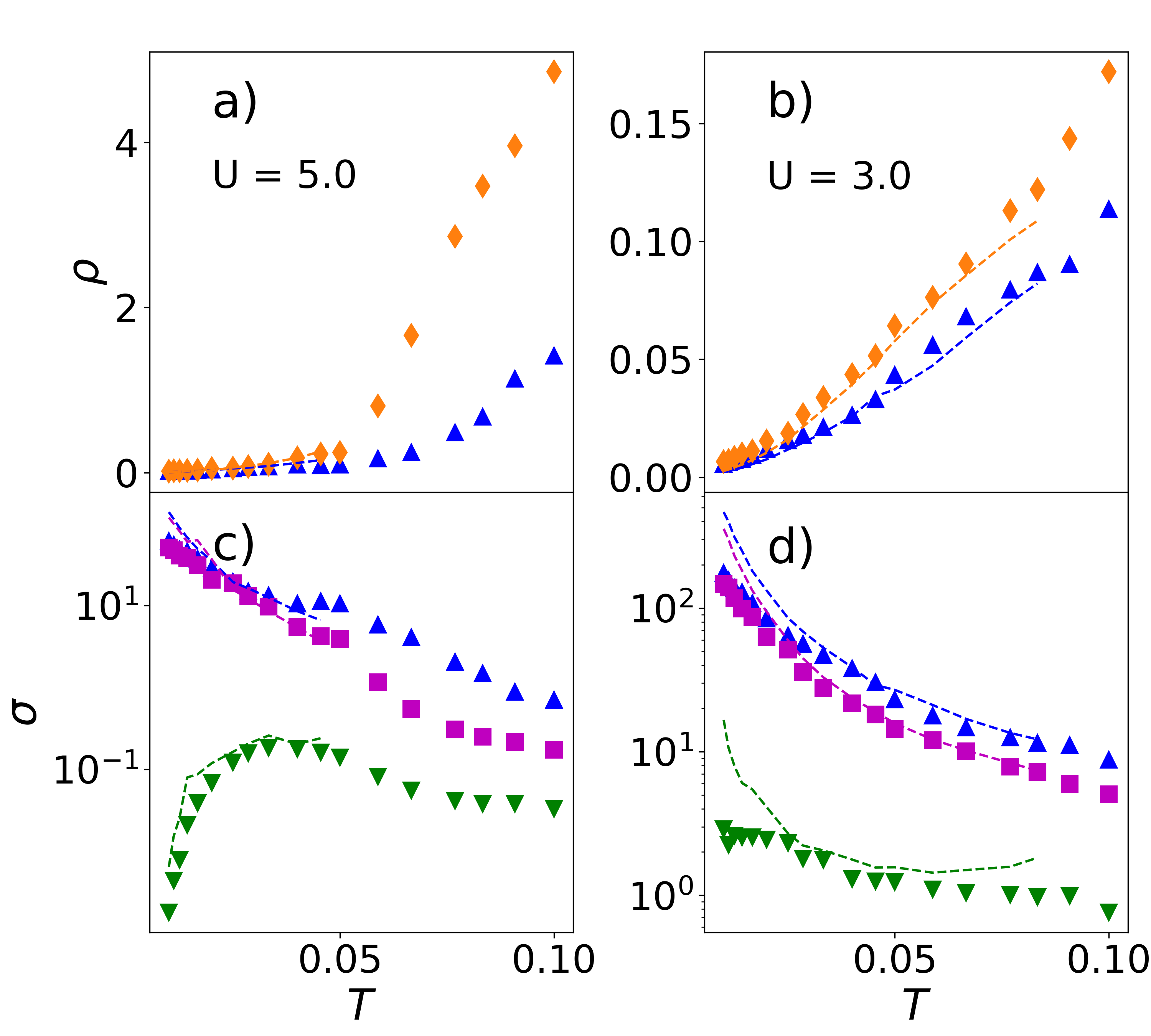}
	\caption{Resistivity and conductivity as a function of temperature. In the upper panels, the resistivity of the 2+1 orbital model (orange) and the 2 orbital model (blue) is plotted for $U = 5.0$, $\Delta_\textrm{CF}=4.0$ in a), and $U = 3.0$, $\Delta_\textrm{CF}=2.5$ in b). The symbols indicate stochastic Maximum Entropy and the dashed lines Padé results. In (c,d) we show the channel resolved conductivities: we plot the contribution of the minority orbitals (green) and the contribution of the majority orbitals 
 (magenta) and compare them to the conductivity of the 2 orbital model (blue).
 }
	\label{fig:im_resistivities}
\end{figure}

So how do these effects manifest in transport? In  Fig.~\ref{fig:im_resistivities}~a) an b) we  show the temperature dependence of the total resistivity of the 2 (blue) and the 2+1 orbital model (orange). Consistent with the behavior seen in the scattering rate, the two calculations behave similarly at low temperatures, but at higher temperatures we see a significantly higher resistivity in the 2+1 orbital model (with the distinction becoming more pronounced at higher interaction strengths). We can attribute this to the stronger correlations due to the onset of the spin crossover. 

The orbitally resolved contributions to conductivity are shown in   Fig.~\ref{fig:im_resistivities} (c, d). The minority orbital contribution in the 2+1 orbital model remains significantly smaller than the majority orbital contribution also at high temperatures.  In terms of the Boltzmann picture, at lower temperatures, scattering in the minority orbital is small but the minority states contribute negligibly due to the low density of electrons there, at high temperature the growth of the density is counteracted by the growth in the scattering rate.

\section{3+2 orbital model}

We now turn to the 3+2 orbital model. We consider two cases, one with a total filling of $N=4$ electrons, as well as a total filling of $N=5$ electrons per atom. Here, we chose an interaction strength $U = 2.4$ and again $J/U = 0.2$, which are reasonable values for many real materials, and yields a correlated metallic state with $Z$ at $\beta=100$ being 0.3 and 0.7 for $N = 4$ and $N = 5$, respectively. 
The atomic criterion for the spin-crossover is $\Delta_\mathrm{CF}=5J=2.4$ for both $N=4$ and $N = 5$, and we chose the crystal field to be slightly larger, $\Delta_\mathrm{CF} = 2.5$ for both fillings.

\begin{figure}[t]
	\includegraphics[width=0.99\columnwidth]{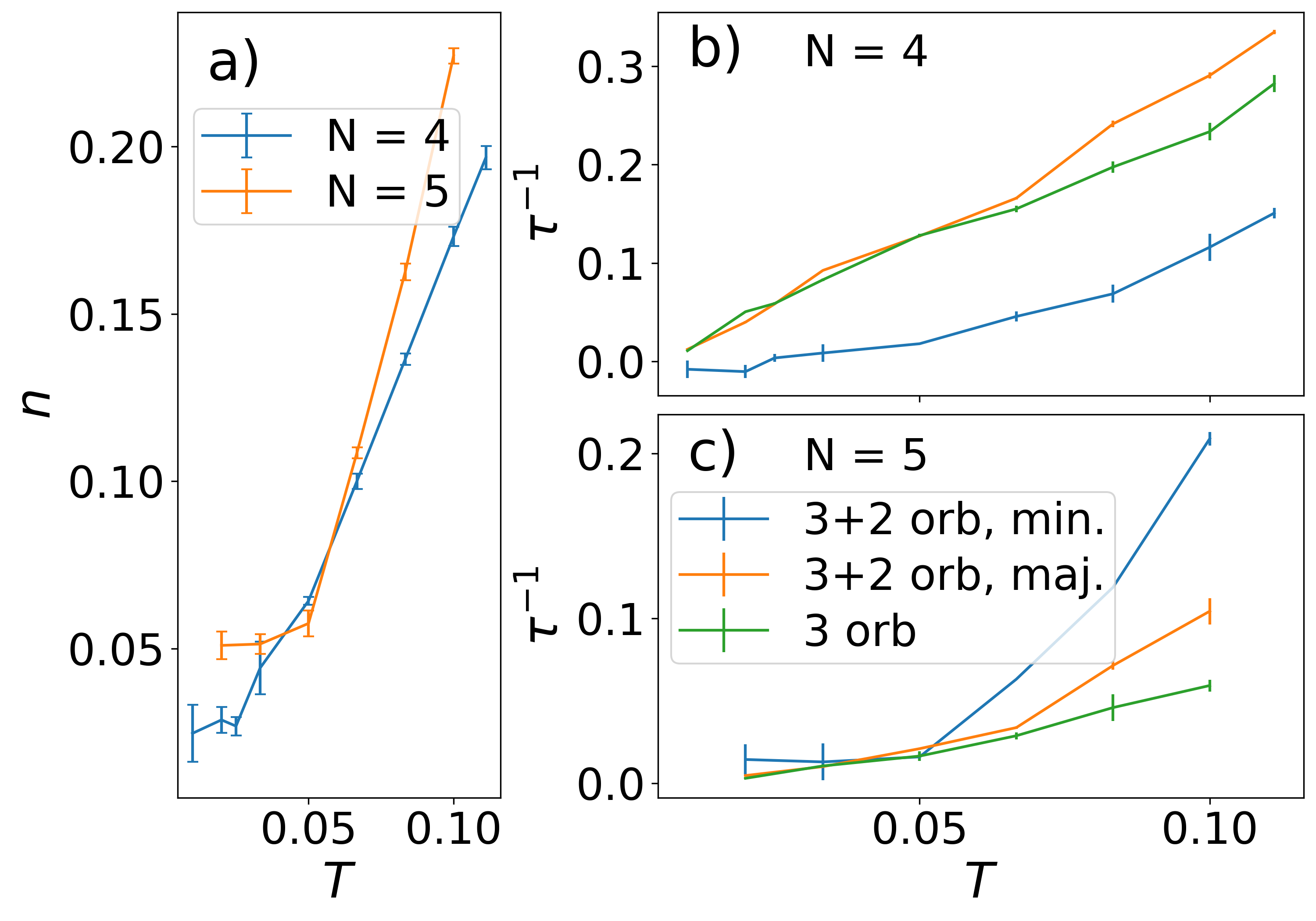}
	\caption{a) Total density in the minority orbitals as a function of temperature for $N = 4$ (blue) and $N = 5$ (orange). The errorbars were obtained by calculating the standard error of the densities over the last 5 (converged) iterations. b) The quasiparticle scattering rate for the case $N=4$ for the minority (blue) and the majority orbital (orange) of the 3+2 orbital model compared to the reference unpolarized 3 orbital model (green). Panel c) shows the same quantities as panel b), but for $N=5$.}
	\label{fig:im_dens_5orb}
\end{figure}
\begin{figure}[t]
\centering
	\includegraphics[width=0.99\columnwidth]{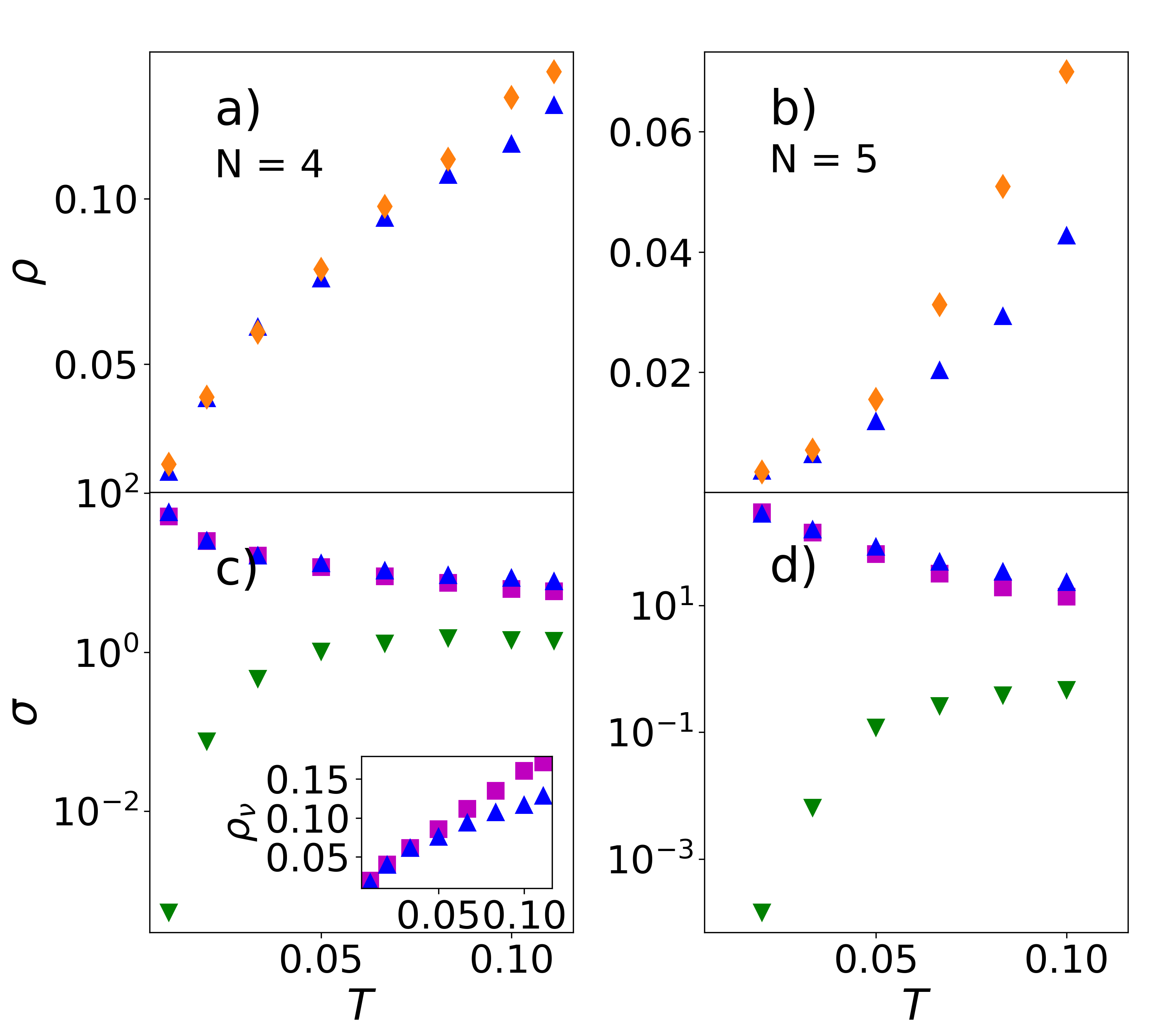}
	\caption{Resistivities (a,b) and channel resolved conductivities (c,d) plotted following the same color code as in Fig.~\ref{fig:im_resistivities}. The inset in the left lower panel shows resistivity of the 3 orbital model (blue triangles) and of the \textit{majority} orbitals of the 3+2 orbital model (violet squares).
 }
	\label{fig:im_resistivity_5orb}
\end{figure}
\begin{figure}[ht]
\centering
	\includegraphics[width=0.99\columnwidth]{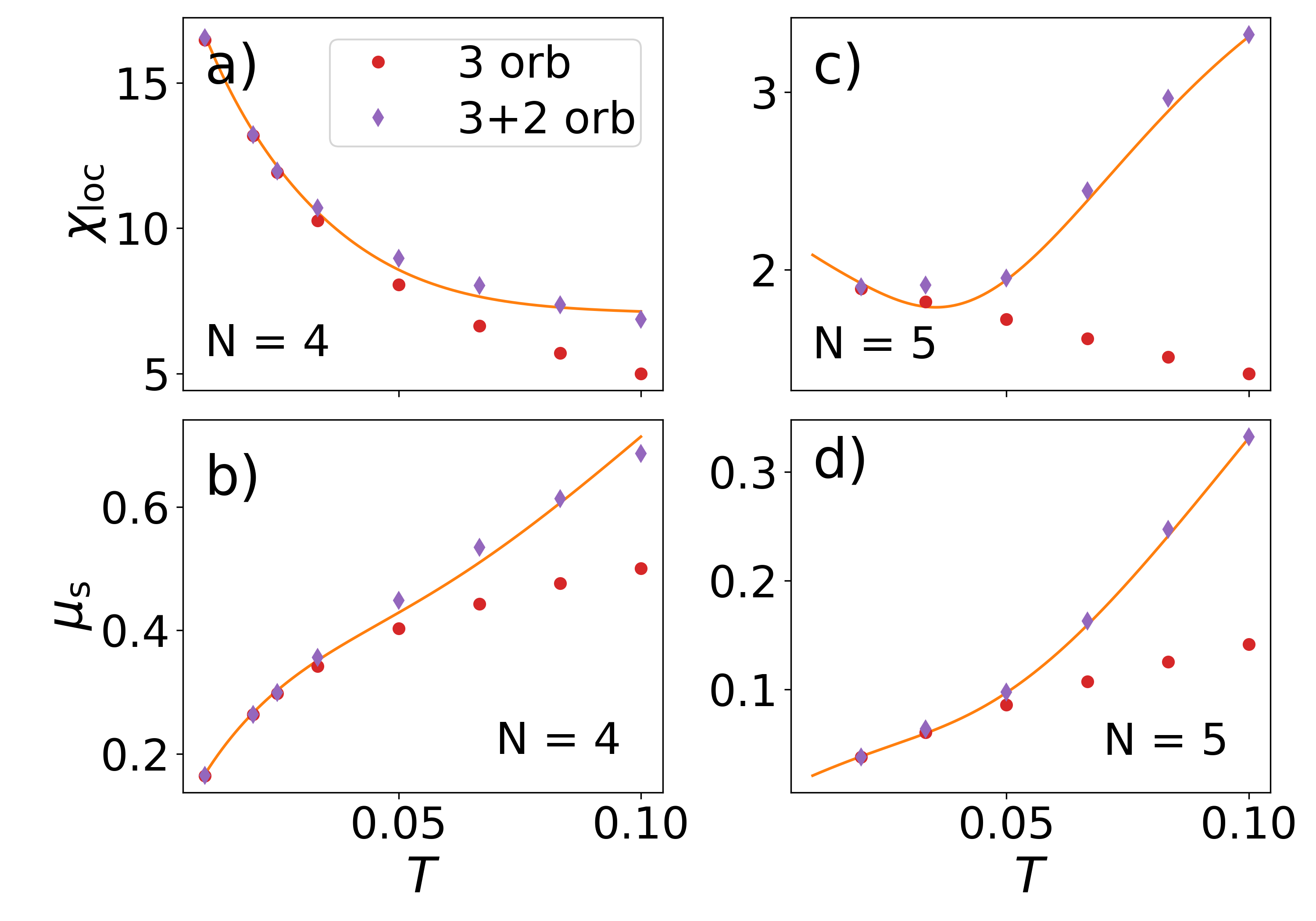}
	\caption{Local magnetic susceptibility and magnetic moment as a function of temperature. In the upper panels (a, c), we show the local magnetic susceptibility of the 3 orbital model (circles) and of the 3+2 orbital model (diamonds) for both $N = 4$ (left) and $N = 5$ (right) calculated using eq. \ref{eq:magsus}, as well as a corresponding fit discussed in the Appendix.  In (c,d) the corresponding magnetic moment $\mu_\mathrm{s}=\chi_\mathrm{loc} T $ is plotted. 
 }
	\label{fig:im_suszept_5orb}
\end{figure}
The temperature dependence of the total minority orbital occupancy $n$ is shown in  Fig.~\ref{fig:im_dens_5orb}~(a). The data shows an increase from 0.01 at low $T$ to about 0.18 for $N=4$ and 0.22 for $N=5$, respectively, at $T=0.1$. 

The scattering rates are shown in Fig.~\ref{fig:im_dens_5orb}~(b,c) for $N=4$ and 5, respectively. The majority electrons behave qualitatively similarly for both cases, their scattering increases faster with temperature (orange lines) compared to the reference 3-orbital case (green lines). The minority electrons on the other hand show an important difference. For $N=4$, the scattering rate does not increase at the same extent as for $N=5$, where for the highest temperatures it even exceeds the one in the majority orbitals. 

Let us look now at the impact on resistivities. We show the results of our calculations in Fig.~\ref{fig:im_resistivity_5orb}, in panel a)  for $N=4$ and in panel b) for $N=5$. We see in both cases a similar effect as before, namely that the total resistivity increases, when additional thermally activated orbitals are taken into consideration, both for $N=4$ and $N=5$. However, for $N=5$ the effect is much stronger, similar in size to the case of $U=5$ in the 2+1 model. Also the origin of this effect is similar, namely that the scattering in the majority orbitals is increased significantly, thus lowering the conductivity there. But at the same time, the additional minority orbital also sees large scattering, and together with the still small electron density, its contribution to conductivity is negligible. This is seen clearly in Fig.~\ref{fig:im_resistivity_5orb}~d), where we show the orbitally-resolved conductivities. The minority orbital, shown with green triangles, even at highest temperatures contributes to the conductivity two orders of magnitude less than the majority orbitals.

This is rather different in the case $N=4$. On the one hand, the effect on the majority orbitals is still similar, meaning that the scattering is increased, which reduces the conductivity. The inset in Fig.~\ref{fig:im_resistivity_5orb}~c) shows this nicely. On the other hand, the contribution of the minority, thermally activated orbitasl, is different. As already shown in Fig.~\ref{fig:im_dens_5orb}~b), the scattering in these orbitals does not increase that much, and stays moderate. Therefore, the additional electronic degrees of freedom there can indeed contribute to conductivity, as shown in the orbitally-resolved conductivities shown in Fig.~\ref{fig:im_resistivity_5orb}~c). Note that we show them on logarithmic scale, spanning four orders of magnitude of $N=4$, compared to seven orders of magnitude for $N=5$. 

As a result, we have two \textit{counteracting} effects; first, the increased resistivity in the majority orbitals, which is obvious from the inset in Fig.~\ref{fig:im_resistivity_5orb}~c), and second the increased conductivity in the minority orbitals. The first effect is quantitatively larger, so there remains an increase in total resistivity, as shown in Fig.~\ref{fig:im_resistivity_5orb}~a).

We also calculated the local magnetic susceptibility, which is shown in Fig.~\ref{fig:im_suszept_5orb}. The fits that follow the same structure (eq. \ref{eq:susfit}) as in the 2+1 orbital model, are again in very good agreement with the data. For $N=5$ we see a non-monotonous behavior as before, just more pronounced (eventually $\chi_\mathrm{loc}$ for the 3+2 model decreases again for very large temperatures). For $N=4$, however, there is no maximum as function of $T$ for considered crystal field. Similarly in the local moment shown in Fig.~\ref{fig:im_suszept_5orb}~b) the distinction between the full calculation and the smaller three orbital calculation is smaller for $N=4$ than for $N=5$. For the 3+2 model and $N=5$, the difference between the low-spin and the high-spin state in terms of total spin is large ($S=1/2$ vs. $S=5/2$), whereas for $N=4$ it is just $S=1$ vs. $S=2$. One should note that the values of magnetic moment are again well below the saturation value $S(S+1)/3$, giving $\mu_\mathrm{s}=2.0$ and $2.92$ for $S=2$ and $S=5/2$, respectively.

\section{Conclusions}

In summary, we studied how the onset of spin crossover caused by thermal activation of carriers to nearly empty correlated orbitals in close proximity impacts the electronic transport and magnetic response of correlated metals. We found rather strong effects, even when the minority orbital carrier density remains small and correspondingly the magnetic moments are still far below the asymptotic high-spin ones. Universally, we found that transfer of electrons to the minority orbitals increases the scattering in the majority ones and leads to a resistivity that is significantly increased. For the case of four electrons in 3+2 orbitals, the minority carriers scatter less and there is a counteracting effect of a parallel conduction channel. However, this effect is less strong, and the total resistivity still increases. 

We envisage our results to be relevant for correlated materials, and ruthenate and rhodate compounds in particular. Signatures of spin-crossover are seen experimentally~\cite{Maeno1997,Perry2006} and in particular in Sr$_2$RhO$_4$ the resistivity is remarkably high reaching ($300$\,$\mu\Omega$cm at room temperature), which is difficult to understand within a $t_{2g}$ picture, as correlations in rhodates are believed to be less important compared to the Hund metal ruthenate compounds~\cite{hund_coupling_jernej}. 

%\section{Data and Code availability}

The major part of this work has been done using the TRIQS package, version 3.1~\cite{triqs}. Codes, scripts, as well as the data shown in this work is publicly available on a permanent repository~\cite{datarep}.

\begin{acknowledgments}

Jernej Mravlje and Markus Aichhorn contributed equally to this work.
We acknowledge support from projects No. P1-0044 and No. J1-2458 of the Slovenian Research Agency (ARIS).
We acknowledge useful discussions with S.~Beck and A.~Georges. Calculations have been performed in parts on the \textit{lcluster} of TU Graz, and on the Vienna Scientific Cluster (VSC5).
\end{acknowledgments}

%\newpage

\appendix

\section{Dependency on Hunds coupling $J$}\label{sec:appB}

In the main text we have determined the parameters of the Hamiltonian, i.e. $U$, $J$, and $\Delta_\textrm{CF}$ such that we are at the point in phase space where the low-spin to high-spin transition, and therefore thermal activation, occurs. In order to understand the influence of variations of parameters, in particular $J$, a bit better, we present here resutls for fixed $U=5.0$ and $\Delta_\textrm{CF}=4.0$, but for $J/U = 0.18$ and $J/U=0.22$.

\begin{figure}[t]
\centering
	\includegraphics[width=0.95\columnwidth]{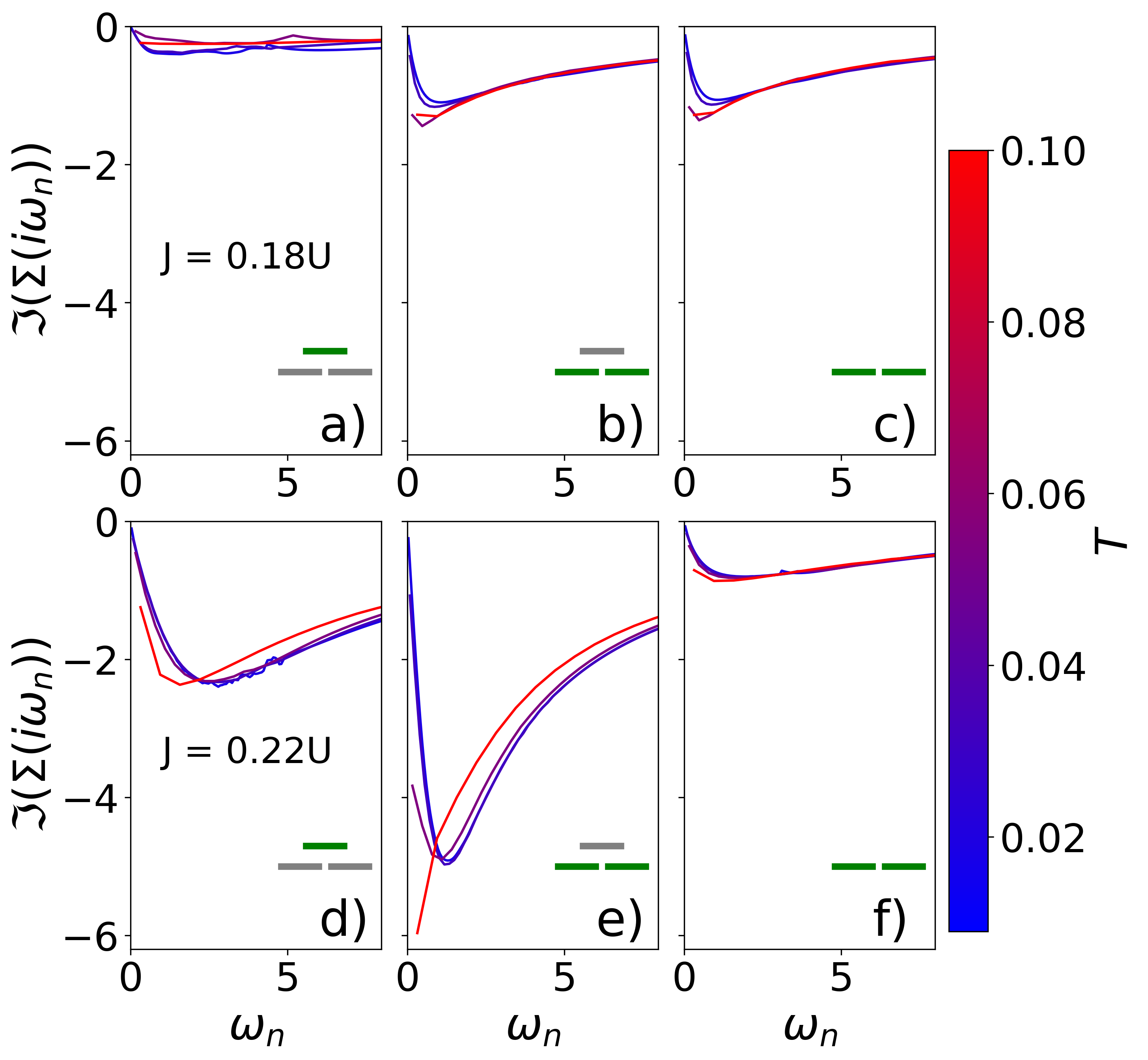}
	\caption{Imaginary part of the self energy of the 2+1 orbital model, shown for $J/U = 0.18$ (a-c) and $J/U = 0.22$ (d-f). Plots a) and d) show the minority orbitals, b) and e) the majority orbitals, and c) and f) show the corresponding 2-orbital model. Different temperatures are indicated by different colors (blue = low temperatures, red = high temperatures). Interaction strength $U = 5.0$ and crystal field $\Delta_\textrm{CF}= 4.0$ are fixed.}
	\label{fig:im_Jdep_Tdep_ver2}
\end{figure}

We see that for a fixed crystal field we quickly leave this point in phase space where low-spin to high-spin transitions occur. If Hunds coupling is too small ($J/U = 0.18$), no thermal activation happens and there is no effect on the resistivities - plots a) to c) in Fig.~\ref{fig:im_Jdep_Tdep_ver2}. If on the other hand the Hunds coupling is too strong ($J/U = 0.22$), the system is already Mott insulating at lowest temperatures - plots d) to f) in Fig.~\ref{fig:im_Jdep_Tdep_ver2}. In both cases, the variation of this state with temperature is small and not relevant for the physical effect that we are interested in. 
%Although this results show a certain fine-tuning aspect of our modelling, we do believe that this transition plays a relevant role in materials such as some ruthenates and rhodates as well as MOFs, as explained in the main text.

\section{Susceptibility for non-degenerate multiplets}\label{sec:appA}

We take the ionic approximation for the case of non-generate multiplets $S$ with energy $E_S$. The general expression for the spin susceptibility then reads
\begin{equation}
\chi \propto \frac{1}{T} \frac{\sum_S g_S (\langle S^2\rangle/3)\exp(- \beta E_S)}  {\sum_S g_S  \exp(- \beta E_S)} \label{eq:VanVleck}, 
\end{equation}
where $g_s$ are the degeneracy factors, $\langle S^2\rangle$ is the size of the spin squared. Above expression is in agreement with early general results of Van Vleck~\cite{vanvleck_original}. We adapt this general form for our case with two multiplets (one at energy $E_S=0$ and one at $E_S=\Delta_\textrm{CF}^{eff}$), and use for fitting
\begin{equation}
\chi = \frac{\gamma \frac{\langle s^2 \rangle/3}{T+T_0} + \tilde{\gamma}\frac{\langle S^2 \rangle/3}{T}e^{-\Delta^{eff}_{CF}/k_BT}}{\gamma + \tilde{\gamma}e^{-\Delta^{eff}_{CF}/k_BT}} \, .
\label{eq:susfit}
\end{equation}
The parameters $T_0$ and $\Delta_\textrm{CF}^{eff}$ are fitting parameters which we obtained using a least squares fitting procedure and have the physical interpretation of the Fermi liquid cutoff, and the effective energy level of the upper orbitals, respectively. 
The $\langle S^2 \rangle$  and $\langle s^2 \rangle$ are taken to be the high- and low-spin expectation values, respectively, and $\gamma$ and $\tilde{\gamma}$ are the degeneracies of the multiplets,
%The other parameters can be calculated ab initio, with $\langle S^2\rangle$ and $\langle s^2\rangle$ being the (high and low) spin expectation values as described above, and $\gamma$ and $\tilde{\gamma}$ being the degeneracies of the multiplets, 
obtained by diagonalising the atomic Hubbard Kanamori Hamiltonian with crystal field. Here it should be noted that there is some ambiguity in the 3+2 orbital model in the case $N = 4$, as the degeneracy of the high spin state $\tilde{\gamma}$ depends on whether the crystal field is included ($\tilde{\gamma} = 10$), or if the degeneracy is calculated using the $S=2$ and $L=2$ degeneracies of 4 electrons in the d-shell, which would lead to $\tilde{\gamma} = 25$. The fits do not change significantly if one uses this definition. 
All parameters of the curves shown in the main text are listed in Tab.~\ref{tab:parameters}.

\begin{table}[ht]
\begin{tabular}{|p{2cm}|p{0.9cm}|p{0.9cm}||p{1cm}|p{1cm}|p{0.6cm}|p{0.6cm}|}
\hline
Model        & $T_0$  & $\Delta_\mathrm{CF}^{eff}$ & $\langle s^2 \rangle/3 $& $\langle S^2 \rangle/3$&$\gamma$ & $\tilde{\gamma}$ \\
\hline
2+1, $U = 5$ & 0.04     & 0.09       &1/4 & 5/4    & 4        & 4   \\
\hline
3+2, $N = 4$ & 0.03   & 0.29    & 2/3           & 4     & 9        & 10  \\
\hline
3+2, $N = 5$ & 0.11    & 0.25  & 1/4      & 35/12         & 6        & 6\\       
\hline
\end{tabular}
\caption{Table of fit parameters $T_0$ and $\Delta_\mathrm{CF}^{eff}$ of Eq. \ref{eq:susfit} for the susceptibilities shown in Fig. \ref{fig:im_susceptibility} and Fig. \ref{fig:im_suszept_5orb}, as well as the according degeneracies $\gamma$ and $\tilde{\gamma}$ and expectation values of the spin operators for the low spin state and high spin state, $\langle s^2 \rangle/3$ and $\langle S^2 \rangle/3$ respectively.}
\label{tab:parameters}
\end{table}

\bibliography{literaturliste}% Produces the bibliography via BibTeX.

\begin{thebibliography}{75}%
\makeatletter
\providecommand \@ifxundefined [1]{%
 \@ifx{#1\undefined}
}%
\providecommand \@ifnum [1]{%
 \ifnum #1\expandafter \@firstoftwo
 \else \expandafter \@secondoftwo
 \fi
}%
\providecommand \@ifx [1]{%
 \ifx #1\expandafter \@firstoftwo
 \else \expandafter \@secondoftwo
 \fi
}%
\providecommand \natexlab [1]{#1}%
\providecommand \enquote  [1]{``#1''}%
\providecommand \bibnamefont  [1]{#1}%
\providecommand \bibfnamefont [1]{#1}%
\providecommand \citenamefont [1]{#1}%
\providecommand \href@noop [0]{\@secondoftwo}%
\providecommand \href [0]{\begingroup \@sanitize@url \@href}%
\providecommand \@href[1]{\@@startlink{#1}\@@href}%
\providecommand \@@href[1]{\endgroup#1\@@endlink}%
\providecommand \@sanitize@url [0]{\catcode `\\12\catcode `\$12\catcode
  `\&12\catcode `\#12\catcode `\^12\catcode `\_12\catcode `\%12\relax}%
\providecommand \@@startlink[1]{}%
\providecommand \@@endlink[0]{}%
\providecommand \url  [0]{\begingroup\@sanitize@url \@url }%
\providecommand \@url [1]{\endgroup\@href {#1}{\urlprefix }}%
\providecommand \urlprefix  [0]{URL }%
\providecommand \Eprint [0]{\href }%
\providecommand \doibase [0]{https://doi.org/}%
\providecommand \selectlanguage [0]{\@gobble}%
\providecommand \bibinfo  [0]{\@secondoftwo}%
\providecommand \bibfield  [0]{\@secondoftwo}%
\providecommand \translation [1]{[#1]}%
\providecommand \BibitemOpen [0]{}%
\providecommand \bibitemStop [0]{}%
\providecommand \bibitemNoStop [0]{.\EOS\space}%
\providecommand \EOS [0]{\spacefactor3000\relax}%
\providecommand \BibitemShut  [1]{\csname bibitem#1\endcsname}%
\let\auto@bib@innerbib\@empty
%</preamble>
\bibitem [{\citenamefont {Landau}(1957)}]{landau}%
  \BibitemOpen
  \bibfield  {author} {\bibinfo {author} {\bibfnamefont {L.}~\bibnamefont
  {Landau}},\ }\bibfield  {title} {\bibinfo {title} {{The theory of a Fermi
  liquid}},\ }\href@noop {} {\bibfield  {journal} {\bibinfo  {journal} {Soviet
  physics JETP}\ }\textbf {\bibinfo {volume} {3}},\ \bibinfo {pages} {920}
  (\bibinfo {year} {1957})}\BibitemShut {NoStop}%
\bibitem [{\citenamefont {Imada}\ \emph {et~al.}(1998)\citenamefont {Imada},
  \citenamefont {Fujimori},\ and\ \citenamefont {Tokura}}]{imada1998}%
  \BibitemOpen
  \bibfield  {author} {\bibinfo {author} {\bibfnamefont {M.}~\bibnamefont
  {Imada}}, \bibinfo {author} {\bibfnamefont {A.}~\bibnamefont {Fujimori}},\
  and\ \bibinfo {author} {\bibfnamefont {Y.}~\bibnamefont {Tokura}},\
  }\bibfield  {title} {\bibinfo {title} {{Metal-insulator transitions}},\
  }\href@noop {} {\bibfield  {journal} {\bibinfo  {journal} {Rev. Mod. Phys.}\
  }\textbf {\bibinfo {volume} {70}},\ \bibinfo {pages} {1039} (\bibinfo {year}
  {1998})}\BibitemShut {NoStop}%
\bibitem [{\citenamefont {L\"ohneysen}\ \emph {et~al.}(2007)\citenamefont
  {L\"ohneysen}, \citenamefont {Rosch}, \citenamefont {Vojta},\ and\
  \citenamefont {W\"olfle}}]{lohneysen2007}%
  \BibitemOpen
  \bibfield  {author} {\bibinfo {author} {\bibfnamefont {H.~v.}\ \bibnamefont
  {L\"ohneysen}}, \bibinfo {author} {\bibfnamefont {A.}~\bibnamefont {Rosch}},
  \bibinfo {author} {\bibfnamefont {M.}~\bibnamefont {Vojta}},\ and\ \bibinfo
  {author} {\bibfnamefont {P.}~\bibnamefont {W\"olfle}},\ }\bibfield  {title}
  {\bibinfo {title} {{Fermi-liquid instabilities at magnetic quantum phase
  transitions}},\ }\href@noop {} {\bibfield  {journal} {\bibinfo  {journal}
  {Rev. Mod. Phys.}\ }\textbf {\bibinfo {volume} {79}},\ \bibinfo {pages}
  {1015} (\bibinfo {year} {2007})}\BibitemShut {NoStop}%
\bibitem [{\citenamefont {Ziman}(2001)}]{ziman}%
  \BibitemOpen
  \bibfield  {author} {\bibinfo {author} {\bibfnamefont {J.~M.}\ \bibnamefont
  {Ziman}},\ }\href@noop {} {\emph {\bibinfo {title} {{Electrons and phonons:
  the theory of transport phenomena in solids}}}}\ (\bibinfo  {publisher}
  {Oxford university press},\ \bibinfo {year} {2001})\BibitemShut {NoStop}%
\bibitem [{\citenamefont {Gunnarsson}\ \emph {et~al.}(2003)\citenamefont
  {Gunnarsson}, \citenamefont {Calandra},\ and\ \citenamefont
  {Han}}]{gunnarsson2003}%
  \BibitemOpen
  \bibfield  {author} {\bibinfo {author} {\bibfnamefont {O.}~\bibnamefont
  {Gunnarsson}}, \bibinfo {author} {\bibfnamefont {M.}~\bibnamefont
  {Calandra}},\ and\ \bibinfo {author} {\bibfnamefont {J.~E.}\ \bibnamefont
  {Han}},\ }\bibfield  {title} {\bibinfo {title} {{Colloquium: Saturation of
  electrical resistivity}},\ }\href@noop {} {\bibfield  {journal} {\bibinfo
  {journal} {Rev. Mod. Phys.}\ }\textbf {\bibinfo {volume} {75}},\ \bibinfo
  {pages} {1085} (\bibinfo {year} {2003})}\BibitemShut {NoStop}%
\bibitem [{\citenamefont {Deng}\ \emph {et~al.}(2013)\citenamefont {Deng},
  \citenamefont {Mravlje}, \citenamefont {\ifmmode~\check{Z}\else
  \v{Z}\fi{}itko}, \citenamefont {Ferrero}, \citenamefont {Kotliar},\ and\
  \citenamefont {Georges}}]{deng2013}%
  \BibitemOpen
  \bibfield  {author} {\bibinfo {author} {\bibfnamefont {X.}~\bibnamefont
  {Deng}}, \bibinfo {author} {\bibfnamefont {J.}~\bibnamefont {Mravlje}},
  \bibinfo {author} {\bibfnamefont {R.}~\bibnamefont {\ifmmode~\check{Z}\else
  \v{Z}\fi{}itko}}, \bibinfo {author} {\bibfnamefont {M.}~\bibnamefont
  {Ferrero}}, \bibinfo {author} {\bibfnamefont {G.}~\bibnamefont {Kotliar}},\
  and\ \bibinfo {author} {\bibfnamefont {A.}~\bibnamefont {Georges}},\
  }\bibfield  {title} {\bibinfo {title} {{How Bad Metals Turn Good:
  Spectroscopic Signatures of Resilient Quasiparticles}},\ }\href@noop {}
  {\bibfield  {journal} {\bibinfo  {journal} {Phys. Rev. Lett.}\ }\textbf
  {\bibinfo {volume} {110}},\ \bibinfo {pages} {086401} (\bibinfo {year}
  {2013})}\BibitemShut {NoStop}%
\bibitem [{\citenamefont {Bruin}\ \emph {et~al.}(2013)\citenamefont {Bruin},
  \citenamefont {Sakai}, \citenamefont {Perry},\ and\ \citenamefont
  {Mackenzie}}]{bruin2013}%
  \BibitemOpen
  \bibfield  {author} {\bibinfo {author} {\bibfnamefont {J.~A.~N.}\
  \bibnamefont {Bruin}}, \bibinfo {author} {\bibfnamefont {H.}~\bibnamefont
  {Sakai}}, \bibinfo {author} {\bibfnamefont {R.~S.}\ \bibnamefont {Perry}},\
  and\ \bibinfo {author} {\bibfnamefont {A.~P.}\ \bibnamefont {Mackenzie}},\
  }\bibfield  {title} {\bibinfo {title} {{Similarity of Scattering Rates in
  Metals Showing T-Linear Resistivity}},\ }\href@noop {} {\bibfield  {journal}
  {\bibinfo  {journal} {Science}\ }\textbf {\bibinfo {volume} {339}},\ \bibinfo
  {pages} {804–807} (\bibinfo {year} {2013})}\BibitemShut {NoStop}%
\bibitem [{\citenamefont {Michon}\ \emph {et~al.}(2023)\citenamefont {Michon},
  \citenamefont {Berthod}, \citenamefont {Rischau}, \citenamefont {Ataei},
  \citenamefont {Chen}, \citenamefont {Komiya}, \citenamefont {Ono},
  \citenamefont {Taillefer}, \citenamefont {van~der Marel},\ and\ \citenamefont
  {Georges}}]{michon2023}%
  \BibitemOpen
  \bibfield  {author} {\bibinfo {author} {\bibfnamefont {B.}~\bibnamefont
  {Michon}}, \bibinfo {author} {\bibfnamefont {C.}~\bibnamefont {Berthod}},
  \bibinfo {author} {\bibfnamefont {C.~W.}\ \bibnamefont {Rischau}}, \bibinfo
  {author} {\bibfnamefont {A.}~\bibnamefont {Ataei}}, \bibinfo {author}
  {\bibfnamefont {L.}~\bibnamefont {Chen}}, \bibinfo {author} {\bibfnamefont
  {S.}~\bibnamefont {Komiya}}, \bibinfo {author} {\bibfnamefont
  {S.}~\bibnamefont {Ono}}, \bibinfo {author} {\bibfnamefont {L.}~\bibnamefont
  {Taillefer}}, \bibinfo {author} {\bibfnamefont {D.}~\bibnamefont {van~der
  Marel}},\ and\ \bibinfo {author} {\bibfnamefont {A.}~\bibnamefont
  {Georges}},\ }\bibfield  {title} {\bibinfo {title} {{Reconciling scaling of
  the optical conductivity of cuprate superconductors with Planckian
  resistivity and specific heat}},\ }\href@noop {} {\bibfield  {journal}
  {\bibinfo  {journal} {Nature Communications}\ }\textbf {\bibinfo {volume}
  {14}},\ \bibinfo {pages} {3033} (\bibinfo {year} {2023})}\BibitemShut
  {NoStop}%
\bibitem [{\citenamefont {Georges}\ and\ \citenamefont
  {Mravlje}(2021)}]{georges2021}%
  \BibitemOpen
  \bibfield  {author} {\bibinfo {author} {\bibfnamefont {A.}~\bibnamefont
  {Georges}}\ and\ \bibinfo {author} {\bibfnamefont {J.}~\bibnamefont
  {Mravlje}},\ }\bibfield  {title} {\bibinfo {title} {{Skewed non-Fermi liquids
  and the Seebeck effect}},\ }\href@noop {} {\bibfield  {journal} {\bibinfo
  {journal} {Phys. Rev. Res.}\ }\textbf {\bibinfo {volume} {3}},\ \bibinfo
  {pages} {043132} (\bibinfo {year} {2021})}\BibitemShut {NoStop}%
\bibitem [{\citenamefont {Abramovitch}\ \emph {et~al.}(2024)\citenamefont
  {Abramovitch}, \citenamefont {Mravlje}, \citenamefont {Zhou}, \citenamefont
  {Georges},\ and\ \citenamefont {Bernardi}}]{abramovitch2024respective}%
  \BibitemOpen
  \bibfield  {author} {\bibinfo {author} {\bibfnamefont {D.~J.}\ \bibnamefont
  {Abramovitch}}, \bibinfo {author} {\bibfnamefont {J.}~\bibnamefont
  {Mravlje}}, \bibinfo {author} {\bibfnamefont {J.-J.}\ \bibnamefont {Zhou}},
  \bibinfo {author} {\bibfnamefont {A.}~\bibnamefont {Georges}},\ and\ \bibinfo
  {author} {\bibfnamefont {M.}~\bibnamefont {Bernardi}},\ }\bibfield  {title}
  {\bibinfo {title} {{Respective Roles of Electron-Phonon and Electron-Electron
  Interactions in the Transport and Quasiparticle Properties of SrVO$_3$}},\
  }\href {https://doi.org/10.1103/PhysRevLett.133.186501} {\bibfield  {journal}
  {\bibinfo  {journal} {Phys. Rev. Lett.}\ }\textbf {\bibinfo {volume} {133}},\
  \bibinfo {pages} {186501} (\bibinfo {year} {2024})}\BibitemShut {NoStop}%
\bibitem [{\citenamefont {Xu}\ \emph {et~al.}(2018)\citenamefont {Xu},
  \citenamefont {Zhang}, \citenamefont {Haule}, \citenamefont {Minar},
  \citenamefont {Wimmer}, \citenamefont {Ebert},\ and\ \citenamefont
  {Cohen}}]{xu2018}%
  \BibitemOpen
  \bibfield  {author} {\bibinfo {author} {\bibfnamefont {J.}~\bibnamefont
  {Xu}}, \bibinfo {author} {\bibfnamefont {P.}~\bibnamefont {Zhang}}, \bibinfo
  {author} {\bibfnamefont {K.}~\bibnamefont {Haule}}, \bibinfo {author}
  {\bibfnamefont {J.}~\bibnamefont {Minar}}, \bibinfo {author} {\bibfnamefont
  {S.}~\bibnamefont {Wimmer}}, \bibinfo {author} {\bibfnamefont
  {H.}~\bibnamefont {Ebert}},\ and\ \bibinfo {author} {\bibfnamefont {R.~E.}\
  \bibnamefont {Cohen}},\ }\bibfield  {title} {\bibinfo {title} {{Thermal
  Conductivity and Electrical Resistivity of Solid Iron at Earth's Core
  Conditions from First Principles}},\ }\href@noop {} {\bibfield  {journal}
  {\bibinfo  {journal} {Phys. Rev. Lett.}\ }\textbf {\bibinfo {volume} {121}},\
  \bibinfo {pages} {096601} (\bibinfo {year} {2018})}\BibitemShut {NoStop}%
\bibitem [{\citenamefont {Pourovskii}\ \emph {et~al.}(2020)\citenamefont
  {Pourovskii}, \citenamefont {Mravlje}, \citenamefont {Pozzo},\ and\
  \citenamefont {Alfè}}]{pourovskii2020}%
  \BibitemOpen
  \bibfield  {author} {\bibinfo {author} {\bibfnamefont {L.~V.}\ \bibnamefont
  {Pourovskii}}, \bibinfo {author} {\bibfnamefont {J.}~\bibnamefont {Mravlje}},
  \bibinfo {author} {\bibfnamefont {M.}~\bibnamefont {Pozzo}},\ and\ \bibinfo
  {author} {\bibfnamefont {D.}~\bibnamefont {Alfè}},\ }\bibfield  {title}
  {\bibinfo {title} {{Electronic correlations and transport in iron at
  Earth’s core conditions}},\ }\href@noop {} {\bibfield  {journal} {\bibinfo
  {journal} {Nature Communications}\ }\textbf {\bibinfo {volume} {11}}
  (\bibinfo {year} {2020})}\BibitemShut {NoStop}%
\bibitem [{\citenamefont {Zhang}\ \emph {et~al.}(2021)\citenamefont {Zhang},
  \citenamefont {Luo}, \citenamefont {Hou}, \citenamefont {Driscoll},
  \citenamefont {Salke}, \citenamefont {Minár}, \citenamefont {Prakapenka},
  \citenamefont {Greenberg}, \citenamefont {Hemley}, \citenamefont {Cohen},\
  and\ \citenamefont {Lin}}]{zhang2021}%
  \BibitemOpen
  \bibfield  {author} {\bibinfo {author} {\bibfnamefont {Y.}~\bibnamefont
  {Zhang}}, \bibinfo {author} {\bibfnamefont {K.}~\bibnamefont {Luo}}, \bibinfo
  {author} {\bibfnamefont {M.}~\bibnamefont {Hou}}, \bibinfo {author}
  {\bibfnamefont {P.}~\bibnamefont {Driscoll}}, \bibinfo {author}
  {\bibfnamefont {N.~P.}\ \bibnamefont {Salke}}, \bibinfo {author}
  {\bibfnamefont {J.}~\bibnamefont {Minár}}, \bibinfo {author} {\bibfnamefont
  {V.~B.}\ \bibnamefont {Prakapenka}}, \bibinfo {author} {\bibfnamefont
  {E.}~\bibnamefont {Greenberg}}, \bibinfo {author} {\bibfnamefont {R.~J.}\
  \bibnamefont {Hemley}}, \bibinfo {author} {\bibfnamefont {R.~E.}\
  \bibnamefont {Cohen}},\ and\ \bibinfo {author} {\bibfnamefont {J.-F.}\
  \bibnamefont {Lin}},\ }\bibfield  {title} {\bibinfo {title} {{Thermal
  conductivity of Fe-Si alloys and thermal stratification in Earth’s core}},\
  }\href@noop {} {\bibfield  {journal} {\bibinfo  {journal} {Proceedings of the
  National Academy of Sciences}\ }\textbf {\bibinfo {volume} {119}},\ \bibinfo
  {pages} {e2119001119} (\bibinfo {year} {2021})}\BibitemShut {NoStop}%
\bibitem [{\citenamefont {Blesio}\ \emph {et~al.}(2025)\citenamefont {Blesio},
  \citenamefont {Pourovskii}, \citenamefont {Aichhorn}, \citenamefont {Pozzo},
  \citenamefont {Alfè},\ and\ \citenamefont {Mravlje}}]{blesio2023influence}%
  \BibitemOpen
  \bibfield  {author} {\bibinfo {author} {\bibfnamefont {G.~G.}\ \bibnamefont
  {Blesio}}, \bibinfo {author} {\bibfnamefont {L.~V.}\ \bibnamefont
  {Pourovskii}}, \bibinfo {author} {\bibfnamefont {M.}~\bibnamefont
  {Aichhorn}}, \bibinfo {author} {\bibfnamefont {M.}~\bibnamefont {Pozzo}},
  \bibinfo {author} {\bibfnamefont {D.}~\bibnamefont {Alfè}},\ and\ \bibinfo
  {author} {\bibfnamefont {J.}~\bibnamefont {Mravlje}},\ }\bibfield  {title}
  {\bibinfo {title} {{Influence of oxygen on electronic correlation and
  transport in iron in the outer Earth's core}},\ }\href@noop {} {\bibfield
  {journal} {\bibinfo  {journal} {Communications Earth and Environment}\
  }\textbf {\bibinfo {volume} {6}},\ \bibinfo {pages} {26} (\bibinfo {year}
  {2025})}\BibitemShut {NoStop}%
\bibitem [{\citenamefont {Abramovitch}\ \emph {et~al.}(2023)\citenamefont
  {Abramovitch}, \citenamefont {Zhou}, \citenamefont {Mravlje}, \citenamefont
  {Georges},\ and\ \citenamefont {Bernardi}}]{abramovitch2023}%
  \BibitemOpen
  \bibfield  {author} {\bibinfo {author} {\bibfnamefont {D.~J.}\ \bibnamefont
  {Abramovitch}}, \bibinfo {author} {\bibfnamefont {J.-J.}\ \bibnamefont
  {Zhou}}, \bibinfo {author} {\bibfnamefont {J.}~\bibnamefont {Mravlje}},
  \bibinfo {author} {\bibfnamefont {A.}~\bibnamefont {Georges}},\ and\ \bibinfo
  {author} {\bibfnamefont {M.}~\bibnamefont {Bernardi}},\ }\bibfield  {title}
  {\bibinfo {title} {{Combining electron-phonon and dynamical mean-field theory
  calculations of correlated materials: Transport in the correlated metal
  ${\mathrm{Sr}}_{2}{\mathrm{RuO}}_{4}$}},\ }\href@noop {} {\bibfield
  {journal} {\bibinfo  {journal} {Phys. Rev. Mater.}\ }\textbf {\bibinfo
  {volume} {7}},\ \bibinfo {pages} {093801} (\bibinfo {year}
  {2023})}\BibitemShut {NoStop}%
\bibitem [{\citenamefont {edited~by P.~G\"utlich}\ and\ \citenamefont
  {Goodwin}(2004)}]{book2004}%
  \BibitemOpen
  \bibfield  {author} {\bibinfo {author} {\bibnamefont {edited~by
  P.~G\"utlich}}\ and\ \bibinfo {author} {\bibfnamefont {H.}~\bibnamefont
  {Goodwin}},\ }\href {https://doi.org/10.1007/b40394-9} {\emph {\bibinfo
  {title} {{Spin Crossover in Transition Metal Compounds I}}}}\ (\bibinfo
  {publisher} {Springer Berlin Heidelberg},\ \bibinfo {year}
  {2004})\BibitemShut {NoStop}%
\bibitem [{\citenamefont {Bari}\ and\ \citenamefont
  {Sivardi\`ere}(1972)}]{lowspin_highspin_propositions1}%
  \BibitemOpen
  \bibfield  {author} {\bibinfo {author} {\bibfnamefont {R.~A.}\ \bibnamefont
  {Bari}}\ and\ \bibinfo {author} {\bibfnamefont {J.}~\bibnamefont
  {Sivardi\`ere}},\ }\bibfield  {title} {\bibinfo {title} {{Low-Spin-High-Spin
  Transitions in Transition-Metal-Ion Compounds}},\ }\href@noop {} {\bibfield
  {journal} {\bibinfo  {journal} {Phys. Rev. B}\ }\textbf {\bibinfo {volume}
  {5}},\ \bibinfo {pages} {4466} (\bibinfo {year} {1972})}\BibitemShut
  {NoStop}%
\bibitem [{\citenamefont {G\"{u}tlich}\ \emph {et~al.}(2000)\citenamefont
  {G\"{u}tlich}, \citenamefont {Garcia},\ and\ \citenamefont
  {Goodwin}}]{Gtlich2000}%
  \BibitemOpen
  \bibfield  {author} {\bibinfo {author} {\bibfnamefont {P.}~\bibnamefont
  {G\"{u}tlich}}, \bibinfo {author} {\bibfnamefont {Y.}~\bibnamefont
  {Garcia}},\ and\ \bibinfo {author} {\bibfnamefont {H.~A.}\ \bibnamefont
  {Goodwin}},\ }\bibfield  {title} {\bibinfo {title} {{Spin crossover phenomena
  in Fe(ii) complexes}},\ }\href {https://doi.org/10.1039/b003504l} {\bibfield
  {journal} {\bibinfo  {journal} {Chemical Society Reviews}\ }\textbf {\bibinfo
  {volume} {29}},\ \bibinfo {pages} {419–427} (\bibinfo {year}
  {2000})}\BibitemShut {NoStop}%
\bibitem [{\citenamefont {Rubio-Giménez}\ \emph {et~al.}(2020)\citenamefont
  {Rubio-Giménez}, \citenamefont {Tatay},\ and\ \citenamefont
  {Martí-Gastaldo}}]{SCO_MOFs}%
  \BibitemOpen
  \bibfield  {author} {\bibinfo {author} {\bibfnamefont {V.}~\bibnamefont
  {Rubio-Giménez}}, \bibinfo {author} {\bibfnamefont {S.}~\bibnamefont
  {Tatay}},\ and\ \bibinfo {author} {\bibfnamefont {C.}~\bibnamefont
  {Martí-Gastaldo}},\ }\bibfield  {title} {\bibinfo {title} {{Electrical
  conductivity and magnetic bistability in metal–organic frameworks and
  coordination polymers: charge transport and spin crossover at the
  nanoscale}},\ }\href@noop {} {\bibfield  {journal} {\bibinfo  {journal}
  {Chem. Soc. Rev.}\ }\textbf {\bibinfo {volume} {49}},\ \bibinfo {pages}
  {5601} (\bibinfo {year} {2020})}\BibitemShut {NoStop}%
\bibitem [{\citenamefont {Martinez-Martinez}\ \emph {et~al.}(2023)\citenamefont
  {Martinez-Martinez}, \citenamefont {Resines-Urien}, \citenamefont
  {Piñeiro-López}, \citenamefont {Fernández-Blanco}, \citenamefont
  {Lorenzo~Mariano}, \citenamefont {Albalad}, \citenamefont {Maspoch},
  \citenamefont {Poloni}, \citenamefont {Rodríguez-Velamazán}, \citenamefont
  {Sañudo}, \citenamefont {Burzurí},\ and\ \citenamefont
  {Sánchez~Costa}}]{SCO_conductivity_2023}%
  \BibitemOpen
  \bibfield  {author} {\bibinfo {author} {\bibfnamefont {A.}~\bibnamefont
  {Martinez-Martinez}}, \bibinfo {author} {\bibfnamefont {E.}~\bibnamefont
  {Resines-Urien}}, \bibinfo {author} {\bibfnamefont {L.}~\bibnamefont
  {Piñeiro-López}}, \bibinfo {author} {\bibfnamefont {A.}~\bibnamefont
  {Fernández-Blanco}}, \bibinfo {author} {\bibfnamefont {A.}~\bibnamefont
  {Lorenzo~Mariano}}, \bibinfo {author} {\bibfnamefont {J.}~\bibnamefont
  {Albalad}}, \bibinfo {author} {\bibfnamefont {D.}~\bibnamefont {Maspoch}},
  \bibinfo {author} {\bibfnamefont {R.}~\bibnamefont {Poloni}}, \bibinfo
  {author} {\bibfnamefont {J.~A.}\ \bibnamefont {Rodríguez-Velamazán}},
  \bibinfo {author} {\bibfnamefont {E.~C.}\ \bibnamefont {Sañudo}}, \bibinfo
  {author} {\bibfnamefont {E.}~\bibnamefont {Burzurí}},\ and\ \bibinfo
  {author} {\bibfnamefont {J.}~\bibnamefont {Sánchez~Costa}},\ }\bibfield
  {title} {\bibinfo {title} {{Spin Crossover-Assisted Modulation of Electron
  Transport in a Single-Crystal 3D Metal–Organic Framework}},\ }\href@noop {}
  {\bibfield  {journal} {\bibinfo  {journal} {Chemistry of Materials}\ }\textbf
  {\bibinfo {volume} {35}},\ \bibinfo {pages} {6012} (\bibinfo {year}
  {2023})}\BibitemShut {NoStop}%
\bibitem [{\citenamefont {Ewald}\ \emph {et~al.}(1964)\citenamefont {Ewald},
  \citenamefont {Martin}, \citenamefont {Ross},\ and\ \citenamefont
  {White}}]{ewald_crossover}%
  \BibitemOpen
  \bibfield  {author} {\bibinfo {author} {\bibfnamefont {A.}~\bibnamefont
  {Ewald}}, \bibinfo {author} {\bibfnamefont {R.}~\bibnamefont {Martin}},
  \bibinfo {author} {\bibfnamefont {I.}~\bibnamefont {Ross}},\ and\ \bibinfo
  {author} {\bibfnamefont {A.}~\bibnamefont {White}},\ }\bibfield  {title}
  {\bibinfo {title} {{Anomalous behaviour at the $^6A_1-^2 T_2$ crossover in
  iron (III) complexes}},\ }\href@noop {} {\bibfield  {journal} {\bibinfo
  {journal} {Proceedings of the Royal Society of London. Series A. Mathematical
  and Physical Sciences}\ }\textbf {\bibinfo {volume} {280}},\ \bibinfo {pages}
  {235} (\bibinfo {year} {1964})}\BibitemShut {NoStop}%
\bibitem [{\citenamefont {Djukic}\ and\ \citenamefont
  {Lemaire}(2009)}]{SCO_CP_conductor}%
  \BibitemOpen
  \bibfield  {author} {\bibinfo {author} {\bibfnamefont {B.}~\bibnamefont
  {Djukic}}\ and\ \bibinfo {author} {\bibfnamefont {M.~T.}\ \bibnamefont
  {Lemaire}},\ }\bibfield  {title} {\bibinfo {title} {{Hybrid Spin-Crossover
  Conductor Exhibiting Unusual Variable-Temperature Electrical Conductivity}},\
  }\href@noop {} {\bibfield  {journal} {\bibinfo  {journal} {Inorganic
  Chemistry}\ }\textbf {\bibinfo {volume} {48}},\ \bibinfo {pages} {10489}
  (\bibinfo {year} {2009})}\BibitemShut {NoStop}%
\bibitem [{\citenamefont {Abbate}\ \emph {et~al.}(1993)\citenamefont {Abbate},
  \citenamefont {Fuggle}, \citenamefont {Fujimori}, \citenamefont {Tjeng},
  \citenamefont {Chen}, \citenamefont {Potze}, \citenamefont {Sawatzky},
  \citenamefont {Eisaki},\ and\ \citenamefont {Uchida}}]{abbate93}%
  \BibitemOpen
  \bibfield  {author} {\bibinfo {author} {\bibfnamefont {M.}~\bibnamefont
  {Abbate}}, \bibinfo {author} {\bibfnamefont {J.~C.}\ \bibnamefont {Fuggle}},
  \bibinfo {author} {\bibfnamefont {A.}~\bibnamefont {Fujimori}}, \bibinfo
  {author} {\bibfnamefont {L.~H.}\ \bibnamefont {Tjeng}}, \bibinfo {author}
  {\bibfnamefont {C.~T.}\ \bibnamefont {Chen}}, \bibinfo {author}
  {\bibfnamefont {R.}~\bibnamefont {Potze}}, \bibinfo {author} {\bibfnamefont
  {G.~A.}\ \bibnamefont {Sawatzky}}, \bibinfo {author} {\bibfnamefont
  {H.}~\bibnamefont {Eisaki}},\ and\ \bibinfo {author} {\bibfnamefont
  {S.}~\bibnamefont {Uchida}},\ }\bibfield  {title} {\bibinfo {title}
  {{Electronic structure and spin-state transition of
  ${\mathrm{LaCoO}}_{3}$}},\ }\href {https://doi.org/10.1103/PhysRevB.47.16124}
  {\bibfield  {journal} {\bibinfo  {journal} {Phys. Rev. B}\ }\textbf {\bibinfo
  {volume} {47}},\ \bibinfo {pages} {16124} (\bibinfo {year}
  {1993})}\BibitemShut {NoStop}%
\bibitem [{\citenamefont {Korotin}\ \emph {et~al.}(1996)\citenamefont
  {Korotin}, \citenamefont {Ezhov}, \citenamefont {Solovyev}, \citenamefont
  {Anisimov}, \citenamefont {Khomskii},\ and\ \citenamefont
  {Sawatzky}}]{korotin96}%
  \BibitemOpen
  \bibfield  {author} {\bibinfo {author} {\bibfnamefont {M.~A.}\ \bibnamefont
  {Korotin}}, \bibinfo {author} {\bibfnamefont {S.~Y.}\ \bibnamefont {Ezhov}},
  \bibinfo {author} {\bibfnamefont {I.~V.}\ \bibnamefont {Solovyev}}, \bibinfo
  {author} {\bibfnamefont {V.~I.}\ \bibnamefont {Anisimov}}, \bibinfo {author}
  {\bibfnamefont {D.~I.}\ \bibnamefont {Khomskii}},\ and\ \bibinfo {author}
  {\bibfnamefont {G.~A.}\ \bibnamefont {Sawatzky}},\ }\bibfield  {title}
  {\bibinfo {title} {{Intermediate-spin state and properties of
  ${\mathrm{LaCoO}}_{3}$}},\ }\href {https://doi.org/10.1103/PhysRevB.54.5309}
  {\bibfield  {journal} {\bibinfo  {journal} {Phys. Rev. B}\ }\textbf {\bibinfo
  {volume} {54}},\ \bibinfo {pages} {5309} (\bibinfo {year}
  {1996})}\BibitemShut {NoStop}%
\bibitem [{\citenamefont {Eder}(2010)}]{eder2010}%
  \BibitemOpen
  \bibfield  {author} {\bibinfo {author} {\bibfnamefont {R.}~\bibnamefont
  {Eder}},\ }\bibfield  {title} {\bibinfo {title} {{Spin-state transition in
  ${\text{LaCoO}}_{3}$ by variational cluster approximation}},\ }\href
  {https://doi.org/10.1103/PhysRevB.81.035101} {\bibfield  {journal} {\bibinfo
  {journal} {Phys. Rev. B}\ }\textbf {\bibinfo {volume} {81}},\ \bibinfo
  {pages} {035101} (\bibinfo {year} {2010})}\BibitemShut {NoStop}%
\bibitem [{\citenamefont {K\ifmmode~\check{r}\else \v{r}\fi{}\'apek}\ \emph
  {et~al.}(2012)\citenamefont {K\ifmmode~\check{r}\else \v{r}\fi{}\'apek},
  \citenamefont {Nov\'ak}, \citenamefont {Kune\ifmmode~\check{s}\else
  \v{s}\fi{}}, \citenamefont {Novoselov}, \citenamefont {Korotin},\ and\
  \citenamefont {Anisimov}}]{krapek2012}%
  \BibitemOpen
  \bibfield  {author} {\bibinfo {author} {\bibfnamefont {V.}~\bibnamefont
  {K\ifmmode~\check{r}\else \v{r}\fi{}\'apek}}, \bibinfo {author}
  {\bibfnamefont {P.}~\bibnamefont {Nov\'ak}}, \bibinfo {author} {\bibfnamefont
  {J.}~\bibnamefont {Kune\ifmmode~\check{s}\else \v{s}\fi{}}}, \bibinfo
  {author} {\bibfnamefont {D.}~\bibnamefont {Novoselov}}, \bibinfo {author}
  {\bibfnamefont {D.~M.}\ \bibnamefont {Korotin}},\ and\ \bibinfo {author}
  {\bibfnamefont {V.~I.}\ \bibnamefont {Anisimov}},\ }\bibfield  {title}
  {\bibinfo {title} {{Spin state transition and covalent bonding in
  LaCoO${}_{3}$}},\ }\href {https://doi.org/10.1103/PhysRevB.86.195104}
  {\bibfield  {journal} {\bibinfo  {journal} {Phys. Rev. B}\ }\textbf {\bibinfo
  {volume} {86}},\ \bibinfo {pages} {195104} (\bibinfo {year}
  {2012})}\BibitemShut {NoStop}%
\bibitem [{\citenamefont {Chakrabarti}\ \emph {et~al.}(2017)\citenamefont
  {Chakrabarti}, \citenamefont {Birol},\ and\ \citenamefont
  {Haule}}]{chakrabarti2017}%
  \BibitemOpen
  \bibfield  {author} {\bibinfo {author} {\bibfnamefont {B.}~\bibnamefont
  {Chakrabarti}}, \bibinfo {author} {\bibfnamefont {T.}~\bibnamefont {Birol}},\
  and\ \bibinfo {author} {\bibfnamefont {K.}~\bibnamefont {Haule}},\ }\bibfield
   {title} {\bibinfo {title} {{Role of entropy and structural parameters in the
  spin-state transition of ${\mathrm{LaCoO}}_{3}$}},\ }\href
  {https://doi.org/10.1103/PhysRevMaterials.1.064403} {\bibfield  {journal}
  {\bibinfo  {journal} {Phys. Rev. Mater.}\ }\textbf {\bibinfo {volume} {1}},\
  \bibinfo {pages} {064403} (\bibinfo {year} {2017})}\BibitemShut {NoStop}%
\bibitem [{\citenamefont {Takegami}\ \emph {et~al.}(2023)\citenamefont
  {Takegami}, \citenamefont {Tanaka}, \citenamefont {Agrestini}, \citenamefont
  {Hu}, \citenamefont {Weinen}, \citenamefont {Rotter}, \citenamefont
  {Sch\"u\ss{}ler-Langeheine}, \citenamefont {Willers}, \citenamefont {Koethe},
  \citenamefont {Lorenz}, \citenamefont {Liao}, \citenamefont {Tsuei},
  \citenamefont {Lin}, \citenamefont {Chen},\ and\ \citenamefont
  {Tjeng}}]{takegami2023}%
  \BibitemOpen
  \bibfield  {author} {\bibinfo {author} {\bibfnamefont {D.}~\bibnamefont
  {Takegami}}, \bibinfo {author} {\bibfnamefont {A.}~\bibnamefont {Tanaka}},
  \bibinfo {author} {\bibfnamefont {S.}~\bibnamefont {Agrestini}}, \bibinfo
  {author} {\bibfnamefont {Z.}~\bibnamefont {Hu}}, \bibinfo {author}
  {\bibfnamefont {J.}~\bibnamefont {Weinen}}, \bibinfo {author} {\bibfnamefont
  {M.}~\bibnamefont {Rotter}}, \bibinfo {author} {\bibfnamefont
  {C.}~\bibnamefont {Sch\"u\ss{}ler-Langeheine}}, \bibinfo {author}
  {\bibfnamefont {T.}~\bibnamefont {Willers}}, \bibinfo {author} {\bibfnamefont
  {T.~C.}\ \bibnamefont {Koethe}}, \bibinfo {author} {\bibfnamefont
  {T.}~\bibnamefont {Lorenz}}, \bibinfo {author} {\bibfnamefont {Y.~F.}\
  \bibnamefont {Liao}}, \bibinfo {author} {\bibfnamefont {K.~D.}\ \bibnamefont
  {Tsuei}}, \bibinfo {author} {\bibfnamefont {H.-J.}\ \bibnamefont {Lin}},
  \bibinfo {author} {\bibfnamefont {C.~T.}\ \bibnamefont {Chen}},\ and\
  \bibinfo {author} {\bibfnamefont {L.~H.}\ \bibnamefont {Tjeng}},\ }\bibfield
  {title} {\bibinfo {title} {{Paramagnetic ${\mathrm{LaCoO}}_{3}$: A Highly
  Inhomogeneous Mixed Spin-State System}},\ }\href
  {https://doi.org/10.1103/PhysRevX.13.011037} {\bibfield  {journal} {\bibinfo
  {journal} {Phys. Rev. X}\ }\textbf {\bibinfo {volume} {13}},\ \bibinfo
  {pages} {011037} (\bibinfo {year} {2023})}\BibitemShut {NoStop}%
\bibitem [{\citenamefont {Wentzcovitch}\ \emph {et~al.}(2010)\citenamefont
  {Wentzcovitch}, \citenamefont {Yu},\ and\ \citenamefont
  {Wu}}]{Wentzcovitch2010}%
  \BibitemOpen
  \bibfield  {author} {\bibinfo {author} {\bibfnamefont {R.~M.}\ \bibnamefont
  {Wentzcovitch}}, \bibinfo {author} {\bibfnamefont {Y.~G.}\ \bibnamefont
  {Yu}},\ and\ \bibinfo {author} {\bibfnamefont {Z.}~\bibnamefont {Wu}},\
  }\bibfield  {title} {\bibinfo {title} {{Thermodynamic Properties and Phase
  Relations in Mantle Minerals Investigated by First Principles Quasiharmonic
  Theory}},\ }\href {https://doi.org/10.2138/rmg.2010.71.4} {\bibfield
  {journal} {\bibinfo  {journal} {Reviews in Mineralogy and Geochemistry}\
  }\textbf {\bibinfo {volume} {71}},\ \bibinfo {pages} {59–98} (\bibinfo
  {year} {2010})}\BibitemShut {NoStop}%
\bibitem [{\citenamefont {Tsuchiya}\ \emph {et~al.}(2006)\citenamefont
  {Tsuchiya}, \citenamefont {Wentzcovitch}, \citenamefont {da~Silva},\ and\
  \citenamefont {de~Gironcoli}}]{tsuchiya06}%
  \BibitemOpen
  \bibfield  {author} {\bibinfo {author} {\bibfnamefont {T.}~\bibnamefont
  {Tsuchiya}}, \bibinfo {author} {\bibfnamefont {R.~M.}\ \bibnamefont
  {Wentzcovitch}}, \bibinfo {author} {\bibfnamefont {C.~R.~S.}\ \bibnamefont
  {da~Silva}},\ and\ \bibinfo {author} {\bibfnamefont {S.}~\bibnamefont
  {de~Gironcoli}},\ }\bibfield  {title} {\bibinfo {title} {{Spin Transition in
  Magnesiow\"ustite in Earth's Lower Mantle}},\ }\href
  {https://doi.org/10.1103/PhysRevLett.96.198501} {\bibfield  {journal}
  {\bibinfo  {journal} {Phys. Rev. Lett.}\ }\textbf {\bibinfo {volume} {96}},\
  \bibinfo {pages} {198501} (\bibinfo {year} {2006})}\BibitemShut {NoStop}%
\bibitem [{\citenamefont {Mackenzie}\ and\ \citenamefont
  {Maeno}(2003)}]{mackenzie03}%
  \BibitemOpen
  \bibfield  {author} {\bibinfo {author} {\bibfnamefont {A.~P.}\ \bibnamefont
  {Mackenzie}}\ and\ \bibinfo {author} {\bibfnamefont {Y.}~\bibnamefont
  {Maeno}},\ }\bibfield  {title} {\bibinfo {title} {{The superconductivity of
  ${\mathrm{Sr}}_{2}{\mathrm{RuO}}_{4}$ and the physics of spin-triplet
  pairing}},\ }\href {https://doi.org/10.1103/RevModPhys.75.657} {\bibfield
  {journal} {\bibinfo  {journal} {Rev. Mod. Phys.}\ }\textbf {\bibinfo {volume}
  {75}},\ \bibinfo {pages} {657} (\bibinfo {year} {2003})}\BibitemShut
  {NoStop}%
\bibitem [{\citenamefont {Perry}\ \emph {et~al.}(2006)\citenamefont {Perry},
  \citenamefont {Baumberger}, \citenamefont {Balicas}, \citenamefont
  {Kikugawa}, \citenamefont {Ingle}, \citenamefont {Rost}, \citenamefont
  {Mercure}, \citenamefont {Maeno}, \citenamefont {Shen},\ and\ \citenamefont
  {Mackenzie}}]{Perry2006}%
  \BibitemOpen
  \bibfield  {author} {\bibinfo {author} {\bibfnamefont {R.~S.}\ \bibnamefont
  {Perry}}, \bibinfo {author} {\bibfnamefont {F.}~\bibnamefont {Baumberger}},
  \bibinfo {author} {\bibfnamefont {L.}~\bibnamefont {Balicas}}, \bibinfo
  {author} {\bibfnamefont {N.}~\bibnamefont {Kikugawa}}, \bibinfo {author}
  {\bibfnamefont {N.~J.~C.}\ \bibnamefont {Ingle}}, \bibinfo {author}
  {\bibfnamefont {A.}~\bibnamefont {Rost}}, \bibinfo {author} {\bibfnamefont
  {J.~F.}\ \bibnamefont {Mercure}}, \bibinfo {author} {\bibfnamefont
  {Y.}~\bibnamefont {Maeno}}, \bibinfo {author} {\bibfnamefont {Z.~X.}\
  \bibnamefont {Shen}},\ and\ \bibinfo {author} {\bibfnamefont {A.~P.}\
  \bibnamefont {Mackenzie}},\ }\bibfield  {title} {\bibinfo {title}
  {{Sr$_2$RhO$_4$: a new, clean correlated electron metal}},\ }\href
  {https://doi.org/10.1088/1367-2630/8/9/175} {\bibfield  {journal} {\bibinfo
  {journal} {New Journal of Physics}\ }\textbf {\bibinfo {volume} {8}},\
  \bibinfo {pages} {175–175} (\bibinfo {year} {2006})}\BibitemShut {NoStop}%
\bibitem [{\citenamefont {Mravlje}\ \emph {et~al.}(2011)\citenamefont
  {Mravlje}, \citenamefont {Aichhorn}, \citenamefont {Miyake}, \citenamefont
  {Haule}, \citenamefont {Kotliar},\ and\ \citenamefont {Georges}}]{mravlje11}%
  \BibitemOpen
  \bibfield  {author} {\bibinfo {author} {\bibfnamefont {J.}~\bibnamefont
  {Mravlje}}, \bibinfo {author} {\bibfnamefont {M.}~\bibnamefont {Aichhorn}},
  \bibinfo {author} {\bibfnamefont {T.}~\bibnamefont {Miyake}}, \bibinfo
  {author} {\bibfnamefont {K.}~\bibnamefont {Haule}}, \bibinfo {author}
  {\bibfnamefont {G.}~\bibnamefont {Kotliar}},\ and\ \bibinfo {author}
  {\bibfnamefont {A.}~\bibnamefont {Georges}},\ }\bibfield  {title} {\bibinfo
  {title} {{Coherence-Incoherence Crossover and the Mass-Renormalization
  Puzzles in ${\mathrm{Sr}}_{2}{\mathrm{RuO}}_{4}$}},\ }\href
  {https://doi.org/10.1103/PhysRevLett.106.096401} {\bibfield  {journal}
  {\bibinfo  {journal} {Phys. Rev. Lett.}\ }\textbf {\bibinfo {volume} {106}},\
  \bibinfo {pages} {096401} (\bibinfo {year} {2011})}\BibitemShut {NoStop}%
\bibitem [{\citenamefont {Behrmann}\ \emph {et~al.}(2012)\citenamefont
  {Behrmann}, \citenamefont {Piefke},\ and\ \citenamefont
  {Lechermann}}]{behrmann12}%
  \BibitemOpen
  \bibfield  {author} {\bibinfo {author} {\bibfnamefont {M.}~\bibnamefont
  {Behrmann}}, \bibinfo {author} {\bibfnamefont {C.}~\bibnamefont {Piefke}},\
  and\ \bibinfo {author} {\bibfnamefont {F.}~\bibnamefont {Lechermann}},\
  }\bibfield  {title} {\bibinfo {title} {{Multiorbital physics in Fermi liquids
  prone to magnetic order}},\ }\href
  {https://doi.org/10.1103/PhysRevB.86.045130} {\bibfield  {journal} {\bibinfo
  {journal} {Phys. Rev. B}\ }\textbf {\bibinfo {volume} {86}},\ \bibinfo
  {pages} {045130} (\bibinfo {year} {2012})}\BibitemShut {NoStop}%
\bibitem [{\citenamefont {Dang}\ \emph
  {et~al.}(2015{\natexlab{a}})\citenamefont {Dang}, \citenamefont {Mravlje},
  \citenamefont {Georges},\ and\ \citenamefont {Millis}}]{dang15}%
  \BibitemOpen
  \bibfield  {author} {\bibinfo {author} {\bibfnamefont {H.~T.}\ \bibnamefont
  {Dang}}, \bibinfo {author} {\bibfnamefont {J.}~\bibnamefont {Mravlje}},
  \bibinfo {author} {\bibfnamefont {A.}~\bibnamefont {Georges}},\ and\ \bibinfo
  {author} {\bibfnamefont {A.~J.}\ \bibnamefont {Millis}},\ }\bibfield  {title}
  {\bibinfo {title} {{Band Structure and Terahertz Optical Conductivity of
  Transition Metal Oxides: Theory and Application to ${\mathrm{CaRuO}}_{3}$}},\
  }\href {https://doi.org/10.1103/PhysRevLett.115.107003} {\bibfield  {journal}
  {\bibinfo  {journal} {Phys. Rev. Lett.}\ }\textbf {\bibinfo {volume} {115}},\
  \bibinfo {pages} {107003} (\bibinfo {year} {2015}{\natexlab{a}})}\BibitemShut
  {NoStop}%
\bibitem [{\citenamefont {Dang}\ \emph
  {et~al.}(2015{\natexlab{b}})\citenamefont {Dang}, \citenamefont {Mravlje},
  \citenamefont {Georges},\ and\ \citenamefont {Millis}}]{dang15prb}%
  \BibitemOpen
  \bibfield  {author} {\bibinfo {author} {\bibfnamefont {H.~T.}\ \bibnamefont
  {Dang}}, \bibinfo {author} {\bibfnamefont {J.}~\bibnamefont {Mravlje}},
  \bibinfo {author} {\bibfnamefont {A.}~\bibnamefont {Georges}},\ and\ \bibinfo
  {author} {\bibfnamefont {A.~J.}\ \bibnamefont {Millis}},\ }\bibfield  {title}
  {\bibinfo {title} {{Electronic correlations, magnetism, and Hund's rule
  coupling in the ruthenium perovskites ${\text{SrRuO}}_{3}$ and
  ${\text{CaRuO}}_{3}$}},\ }\href {https://doi.org/10.1103/PhysRevB.91.195149}
  {\bibfield  {journal} {\bibinfo  {journal} {Phys. Rev. B}\ }\textbf {\bibinfo
  {volume} {91}},\ \bibinfo {pages} {195149} (\bibinfo {year}
  {2015}{\natexlab{b}})}\BibitemShut {NoStop}%
\bibitem [{\citenamefont {Mravlje}\ and\ \citenamefont
  {Georges}(2016)}]{mravlje16}%
  \BibitemOpen
  \bibfield  {author} {\bibinfo {author} {\bibfnamefont {J.}~\bibnamefont
  {Mravlje}}\ and\ \bibinfo {author} {\bibfnamefont {A.}~\bibnamefont
  {Georges}},\ }\bibfield  {title} {\bibinfo {title} {{Thermopower and Entropy:
  Lessons from ${\mathrm{Sr}}_{2}{\mathrm{RuO}}_{4}$}},\ }\href
  {https://doi.org/10.1103/PhysRevLett.117.036401} {\bibfield  {journal}
  {\bibinfo  {journal} {Phys. Rev. Lett.}\ }\textbf {\bibinfo {volume} {117}},\
  \bibinfo {pages} {036401} (\bibinfo {year} {2016})}\BibitemShut {NoStop}%
\bibitem [{\citenamefont {Han}\ \emph {et~al.}(2016)\citenamefont {Han},
  \citenamefont {Dang},\ and\ \citenamefont {Millis}}]{han16}%
  \BibitemOpen
  \bibfield  {author} {\bibinfo {author} {\bibfnamefont {Q.}~\bibnamefont
  {Han}}, \bibinfo {author} {\bibfnamefont {H.~T.}\ \bibnamefont {Dang}},\ and\
  \bibinfo {author} {\bibfnamefont {A.~J.}\ \bibnamefont {Millis}},\ }\bibfield
   {title} {\bibinfo {title} {{Ferromagnetism and correlation strength in cubic
  barium ruthenate in comparison to strontium and calcium ruthenate: A
  dynamical mean-field study}},\ }\href
  {https://doi.org/10.1103/PhysRevB.93.155103} {\bibfield  {journal} {\bibinfo
  {journal} {Phys. Rev. B}\ }\textbf {\bibinfo {volume} {93}},\ \bibinfo
  {pages} {155103} (\bibinfo {year} {2016})}\BibitemShut {NoStop}%
\bibitem [{\citenamefont {Dasari}\ \emph {et~al.}(2016)\citenamefont {Dasari},
  \citenamefont {Yamijala}, \citenamefont {Jain}, \citenamefont {Dasgupta},
  \citenamefont {Moreno}, \citenamefont {Jarrell},\ and\ \citenamefont
  {Vidhyadhiraja}}]{dasari16}%
  \BibitemOpen
  \bibfield  {author} {\bibinfo {author} {\bibfnamefont {N.}~\bibnamefont
  {Dasari}}, \bibinfo {author} {\bibfnamefont {S.~R. K. C.~S.}\ \bibnamefont
  {Yamijala}}, \bibinfo {author} {\bibfnamefont {M.}~\bibnamefont {Jain}},
  \bibinfo {author} {\bibfnamefont {T.~S.}\ \bibnamefont {Dasgupta}}, \bibinfo
  {author} {\bibfnamefont {J.}~\bibnamefont {Moreno}}, \bibinfo {author}
  {\bibfnamefont {M.}~\bibnamefont {Jarrell}},\ and\ \bibinfo {author}
  {\bibfnamefont {N.~S.}\ \bibnamefont {Vidhyadhiraja}},\ }\bibfield  {title}
  {\bibinfo {title} {{First-principles investigation of cubic
  ${\mathrm{BaRuO}}_{3}$: A Hund's metal}},\ }\href
  {https://doi.org/10.1103/PhysRevB.94.085143} {\bibfield  {journal} {\bibinfo
  {journal} {Phys. Rev. B}\ }\textbf {\bibinfo {volume} {94}},\ \bibinfo
  {pages} {085143} (\bibinfo {year} {2016})}\BibitemShut {NoStop}%
\bibitem [{\citenamefont {Zhang}\ \emph {et~al.}(2016)\citenamefont {Zhang},
  \citenamefont {Gorelov}, \citenamefont {Sarvestani},\ and\ \citenamefont
  {Pavarini}}]{zhang16}%
  \BibitemOpen
  \bibfield  {author} {\bibinfo {author} {\bibfnamefont {G.}~\bibnamefont
  {Zhang}}, \bibinfo {author} {\bibfnamefont {E.}~\bibnamefont {Gorelov}},
  \bibinfo {author} {\bibfnamefont {E.}~\bibnamefont {Sarvestani}},\ and\
  \bibinfo {author} {\bibfnamefont {E.}~\bibnamefont {Pavarini}},\ }\bibfield
  {title} {\bibinfo {title} {{Fermi Surface of
  ${\mathrm{Sr}}_{2}{\mathrm{RuO}}_{4}$: Spin-Orbit and Anisotropic Coulomb
  Interaction Effects}},\ }\href
  {https://doi.org/10.1103/PhysRevLett.116.106402} {\bibfield  {journal}
  {\bibinfo  {journal} {Phys. Rev. Lett.}\ }\textbf {\bibinfo {volume} {116}},\
  \bibinfo {pages} {106402} (\bibinfo {year} {2016})}\BibitemShut {NoStop}%
\bibitem [{\citenamefont {Kondo}\ \emph {et~al.}(2016)\citenamefont {Kondo},
  \citenamefont {Ochi}, \citenamefont {Nakayama}, \citenamefont {Taniguchi},
  \citenamefont {Akebi}, \citenamefont {Kuroda}, \citenamefont {Arita},
  \citenamefont {Sakai}, \citenamefont {Namatame}, \citenamefont {Taniguchi},
  \citenamefont {Maeno}, \citenamefont {Arita},\ and\ \citenamefont
  {Shin}}]{kondo16}%
  \BibitemOpen
  \bibfield  {author} {\bibinfo {author} {\bibfnamefont {T.}~\bibnamefont
  {Kondo}}, \bibinfo {author} {\bibfnamefont {M.}~\bibnamefont {Ochi}},
  \bibinfo {author} {\bibfnamefont {M.}~\bibnamefont {Nakayama}}, \bibinfo
  {author} {\bibfnamefont {H.}~\bibnamefont {Taniguchi}}, \bibinfo {author}
  {\bibfnamefont {S.}~\bibnamefont {Akebi}}, \bibinfo {author} {\bibfnamefont
  {K.}~\bibnamefont {Kuroda}}, \bibinfo {author} {\bibfnamefont
  {M.}~\bibnamefont {Arita}}, \bibinfo {author} {\bibfnamefont
  {S.}~\bibnamefont {Sakai}}, \bibinfo {author} {\bibfnamefont
  {H.}~\bibnamefont {Namatame}}, \bibinfo {author} {\bibfnamefont
  {M.}~\bibnamefont {Taniguchi}}, \bibinfo {author} {\bibfnamefont
  {Y.}~\bibnamefont {Maeno}}, \bibinfo {author} {\bibfnamefont
  {R.}~\bibnamefont {Arita}},\ and\ \bibinfo {author} {\bibfnamefont
  {S.}~\bibnamefont {Shin}},\ }\bibfield  {title} {\bibinfo {title}
  {{Orbital-Dependent Band Narrowing Revealed in an Extremely Correlated Hund's
  Metal Emerging on the Topmost Layer of
  ${\mathrm{Sr}}_{2}{\mathrm{RuO}}_{4}$}},\ }\href
  {https://doi.org/10.1103/PhysRevLett.117.247001} {\bibfield  {journal}
  {\bibinfo  {journal} {Phys. Rev. Lett.}\ }\textbf {\bibinfo {volume} {117}},\
  \bibinfo {pages} {247001} (\bibinfo {year} {2016})}\BibitemShut {NoStop}%
\bibitem [{\citenamefont {Kim}\ \emph {et~al.}(2018)\citenamefont {Kim},
  \citenamefont {Mravlje}, \citenamefont {Ferrero}, \citenamefont {Parcollet},\
  and\ \citenamefont {Georges}}]{kim18}%
  \BibitemOpen
  \bibfield  {author} {\bibinfo {author} {\bibfnamefont {M.}~\bibnamefont
  {Kim}}, \bibinfo {author} {\bibfnamefont {J.}~\bibnamefont {Mravlje}},
  \bibinfo {author} {\bibfnamefont {M.}~\bibnamefont {Ferrero}}, \bibinfo
  {author} {\bibfnamefont {O.}~\bibnamefont {Parcollet}},\ and\ \bibinfo
  {author} {\bibfnamefont {A.}~\bibnamefont {Georges}},\ }\bibfield  {title}
  {\bibinfo {title} {{Spin-Orbit Coupling and Electronic Correlations in
  ${\mathrm{Sr}}_{2}{\mathrm{RuO}}_{4}$}},\ }\href
  {https://doi.org/10.1103/PhysRevLett.120.126401} {\bibfield  {journal}
  {\bibinfo  {journal} {Phys. Rev. Lett.}\ }\textbf {\bibinfo {volume} {120}},\
  \bibinfo {pages} {126401} (\bibinfo {year} {2018})}\BibitemShut {NoStop}%
\bibitem [{\citenamefont {Facio}\ \emph {et~al.}(2018)\citenamefont {Facio},
  \citenamefont {Mravlje}, \citenamefont {Pourovskii}, \citenamefont
  {Cornaglia},\ and\ \citenamefont {Vildosola}}]{facio18}%
  \BibitemOpen
  \bibfield  {author} {\bibinfo {author} {\bibfnamefont {J.~I.}\ \bibnamefont
  {Facio}}, \bibinfo {author} {\bibfnamefont {J.}~\bibnamefont {Mravlje}},
  \bibinfo {author} {\bibfnamefont {L.}~\bibnamefont {Pourovskii}}, \bibinfo
  {author} {\bibfnamefont {P.~S.}\ \bibnamefont {Cornaglia}},\ and\ \bibinfo
  {author} {\bibfnamefont {V.}~\bibnamefont {Vildosola}},\ }\bibfield  {title}
  {\bibinfo {title} {{Spin-orbit and anisotropic strain effects on the
  electronic correlations in ${\mathrm{Sr}}_{2}{\mathrm{RuO}}_{4}$}},\ }\href
  {https://doi.org/10.1103/PhysRevB.98.085121} {\bibfield  {journal} {\bibinfo
  {journal} {Phys. Rev. B}\ }\textbf {\bibinfo {volume} {98}},\ \bibinfo
  {pages} {085121} (\bibinfo {year} {2018})}\BibitemShut {NoStop}%
\bibitem [{\citenamefont {Tamai}\ \emph {et~al.}(2019)\citenamefont {Tamai},
  \citenamefont {Zingl}, \citenamefont {Rozbicki}, \citenamefont {Cappelli},
  \citenamefont {Ricc\`o}, \citenamefont {de~la Torre}, \citenamefont
  {McKeown~Walker}, \citenamefont {Bruno}, \citenamefont {King}, \citenamefont
  {Meevasana}, \citenamefont {Shi}, \citenamefont
  {Radovi\ifmmode~\acute{c}\else \'{c}\fi{}}, \citenamefont {Plumb},
  \citenamefont {Gibbs}, \citenamefont {Mackenzie}, \citenamefont {Berthod},
  \citenamefont {Strand}, \citenamefont {Kim}, \citenamefont {Georges},\ and\
  \citenamefont {Baumberger}}]{tamai19}%
  \BibitemOpen
  \bibfield  {author} {\bibinfo {author} {\bibfnamefont {A.}~\bibnamefont
  {Tamai}}, \bibinfo {author} {\bibfnamefont {M.}~\bibnamefont {Zingl}},
  \bibinfo {author} {\bibfnamefont {E.}~\bibnamefont {Rozbicki}}, \bibinfo
  {author} {\bibfnamefont {E.}~\bibnamefont {Cappelli}}, \bibinfo {author}
  {\bibfnamefont {S.}~\bibnamefont {Ricc\`o}}, \bibinfo {author} {\bibfnamefont
  {A.}~\bibnamefont {de~la Torre}}, \bibinfo {author} {\bibfnamefont
  {S.}~\bibnamefont {McKeown~Walker}}, \bibinfo {author} {\bibfnamefont
  {F.~Y.}\ \bibnamefont {Bruno}}, \bibinfo {author} {\bibfnamefont {P.~D.~C.}\
  \bibnamefont {King}}, \bibinfo {author} {\bibfnamefont {W.}~\bibnamefont
  {Meevasana}}, \bibinfo {author} {\bibfnamefont {M.}~\bibnamefont {Shi}},
  \bibinfo {author} {\bibfnamefont {M.}~\bibnamefont
  {Radovi\ifmmode~\acute{c}\else \'{c}\fi{}}}, \bibinfo {author} {\bibfnamefont
  {N.~C.}\ \bibnamefont {Plumb}}, \bibinfo {author} {\bibfnamefont {A.~S.}\
  \bibnamefont {Gibbs}}, \bibinfo {author} {\bibfnamefont {A.~P.}\ \bibnamefont
  {Mackenzie}}, \bibinfo {author} {\bibfnamefont {C.}~\bibnamefont {Berthod}},
  \bibinfo {author} {\bibfnamefont {H.~U.~R.}\ \bibnamefont {Strand}}, \bibinfo
  {author} {\bibfnamefont {M.}~\bibnamefont {Kim}}, \bibinfo {author}
  {\bibfnamefont {A.}~\bibnamefont {Georges}},\ and\ \bibinfo {author}
  {\bibfnamefont {F.}~\bibnamefont {Baumberger}},\ }\bibfield  {title}
  {\bibinfo {title} {{High-Resolution Photoemission on
  ${\mathrm{Sr}}_{2}{\mathrm{RuO}}_{4}$ Reveals Correlation-Enhanced Effective
  Spin-Orbit Coupling and Dominantly Local Self-Energies}},\ }\href
  {https://doi.org/10.1103/PhysRevX.9.021048} {\bibfield  {journal} {\bibinfo
  {journal} {Phys. Rev. X}\ }\textbf {\bibinfo {volume} {9}},\ \bibinfo {pages}
  {021048} (\bibinfo {year} {2019})}\BibitemShut {NoStop}%
\bibitem [{\citenamefont {Strand}\ \emph {et~al.}(2019)\citenamefont {Strand},
  \citenamefont {Zingl}, \citenamefont {Wentzell}, \citenamefont {Parcollet},\
  and\ \citenamefont {Georges}}]{strand19}%
  \BibitemOpen
  \bibfield  {author} {\bibinfo {author} {\bibfnamefont {H.~U.~R.}\
  \bibnamefont {Strand}}, \bibinfo {author} {\bibfnamefont {M.}~\bibnamefont
  {Zingl}}, \bibinfo {author} {\bibfnamefont {N.}~\bibnamefont {Wentzell}},
  \bibinfo {author} {\bibfnamefont {O.}~\bibnamefont {Parcollet}},\ and\
  \bibinfo {author} {\bibfnamefont {A.}~\bibnamefont {Georges}},\ }\bibfield
  {title} {\bibinfo {title} {{Magnetic response of
  ${\mathrm{Sr}}_{2}{\mathrm{RuO}}_{4}$: Quasi-local spin fluctuations due to
  Hund's coupling}},\ }\href {https://doi.org/10.1103/PhysRevB.100.125120}
  {\bibfield  {journal} {\bibinfo  {journal} {Phys. Rev. B}\ }\textbf {\bibinfo
  {volume} {100}},\ \bibinfo {pages} {125120} (\bibinfo {year}
  {2019})}\BibitemShut {NoStop}%
\bibitem [{\citenamefont {Sarvestani}\ \emph {et~al.}(2018)\citenamefont
  {Sarvestani}, \citenamefont {Zhang}, \citenamefont {Gorelov},\ and\
  \citenamefont {Pavarini}}]{sarvestani19}%
  \BibitemOpen
  \bibfield  {author} {\bibinfo {author} {\bibfnamefont {E.}~\bibnamefont
  {Sarvestani}}, \bibinfo {author} {\bibfnamefont {G.}~\bibnamefont {Zhang}},
  \bibinfo {author} {\bibfnamefont {E.}~\bibnamefont {Gorelov}},\ and\ \bibinfo
  {author} {\bibfnamefont {E.}~\bibnamefont {Pavarini}},\ }\bibfield  {title}
  {\bibinfo {title} {{Effective masses, lifetimes, and optical conductivity in
  ${\mathrm{Sr}}_{2}{\mathrm{RuO}}_{4}$ and
  ${\mathrm{Sr}}_{3}{\mathrm{Ru}}_{2}{\mathrm{O}}_{7}$: Interplay of
  spin-orbit, crystal-field, and Coulomb tetragonal tensor interactions}},\
  }\href {https://doi.org/10.1103/PhysRevB.97.085141} {\bibfield  {journal}
  {\bibinfo  {journal} {Phys. Rev. B}\ }\textbf {\bibinfo {volume} {97}},\
  \bibinfo {pages} {085141} (\bibinfo {year} {2018})}\BibitemShut {NoStop}%
\bibitem [{\citenamefont {Kugler}\ \emph {et~al.}(2020)\citenamefont {Kugler},
  \citenamefont {Zingl}, \citenamefont {Strand}, \citenamefont {Lee},
  \citenamefont {von Delft},\ and\ \citenamefont {Georges}}]{kugler20}%
  \BibitemOpen
  \bibfield  {author} {\bibinfo {author} {\bibfnamefont {F.~B.}\ \bibnamefont
  {Kugler}}, \bibinfo {author} {\bibfnamefont {M.}~\bibnamefont {Zingl}},
  \bibinfo {author} {\bibfnamefont {H.~U.~R.}\ \bibnamefont {Strand}}, \bibinfo
  {author} {\bibfnamefont {S.-S.~B.}\ \bibnamefont {Lee}}, \bibinfo {author}
  {\bibfnamefont {J.}~\bibnamefont {von Delft}},\ and\ \bibinfo {author}
  {\bibfnamefont {A.}~\bibnamefont {Georges}},\ }\bibfield  {title} {\bibinfo
  {title} {{Strongly Correlated Materials from a Numerical Renormalization
  Group Perspective: How the Fermi-Liquid State of
  ${\mathrm{Sr}}_{2}{\mathrm{RuO}}_{4}$ Emerges}},\ }\href
  {https://doi.org/10.1103/PhysRevLett.124.016401} {\bibfield  {journal}
  {\bibinfo  {journal} {Phys. Rev. Lett.}\ }\textbf {\bibinfo {volume} {124}},\
  \bibinfo {pages} {016401} (\bibinfo {year} {2020})}\BibitemShut {NoStop}%
\bibitem [{\citenamefont {Linden}\ \emph {et~al.}(2020)\citenamefont {Linden},
  \citenamefont {Zingl}, \citenamefont {Hubig}, \citenamefont {Parcollet},\
  and\ \citenamefont {Schollw\"ock}}]{linden20}%
  \BibitemOpen
  \bibfield  {author} {\bibinfo {author} {\bibfnamefont {N.-O.}\ \bibnamefont
  {Linden}}, \bibinfo {author} {\bibfnamefont {M.}~\bibnamefont {Zingl}},
  \bibinfo {author} {\bibfnamefont {C.}~\bibnamefont {Hubig}}, \bibinfo
  {author} {\bibfnamefont {O.}~\bibnamefont {Parcollet}},\ and\ \bibinfo
  {author} {\bibfnamefont {U.}~\bibnamefont {Schollw\"ock}},\ }\bibfield
  {title} {\bibinfo {title} {{Imaginary-time matrix product state impurity
  solver in a real material calculation: Spin-orbit coupling in
  $\mathrm{Sr}{}_{2}\mathrm{RuO}{}_{4}$}},\ }\href
  {https://doi.org/10.1103/PhysRevB.101.041101} {\bibfield  {journal} {\bibinfo
   {journal} {Phys. Rev. B}\ }\textbf {\bibinfo {volume} {101}},\ \bibinfo
  {pages} {041101} (\bibinfo {year} {2020})}\BibitemShut {NoStop}%
\bibitem [{\citenamefont {Lee}\ \emph {et~al.}(2021)\citenamefont {Lee},
  \citenamefont {Kim},\ and\ \citenamefont {Go}}]{lee21}%
  \BibitemOpen
  \bibfield  {author} {\bibinfo {author} {\bibfnamefont {H.~J.}\ \bibnamefont
  {Lee}}, \bibinfo {author} {\bibfnamefont {C.~H.}\ \bibnamefont {Kim}},\ and\
  \bibinfo {author} {\bibfnamefont {A.}~\bibnamefont {Go}},\ }\bibfield
  {title} {\bibinfo {title} {{Hund's metallicity enhanced by a van Hove
  singularity in cubic perovskite systems}},\ }\href
  {https://doi.org/10.1103/PhysRevB.104.165138} {\bibfield  {journal} {\bibinfo
   {journal} {Phys. Rev. B}\ }\textbf {\bibinfo {volume} {104}},\ \bibinfo
  {pages} {165138} (\bibinfo {year} {2021})}\BibitemShut {NoStop}%
\bibitem [{\citenamefont {K\"aser}\ \emph {et~al.}(2022)\citenamefont
  {K\"aser}, \citenamefont {Strand}, \citenamefont {Wentzell}, \citenamefont
  {Georges}, \citenamefont {Parcollet},\ and\ \citenamefont
  {Hansmann}}]{kaeser22}%
  \BibitemOpen
  \bibfield  {author} {\bibinfo {author} {\bibfnamefont {S.}~\bibnamefont
  {K\"aser}}, \bibinfo {author} {\bibfnamefont {H.~U.~R.}\ \bibnamefont
  {Strand}}, \bibinfo {author} {\bibfnamefont {N.}~\bibnamefont {Wentzell}},
  \bibinfo {author} {\bibfnamefont {A.}~\bibnamefont {Georges}}, \bibinfo
  {author} {\bibfnamefont {O.}~\bibnamefont {Parcollet}},\ and\ \bibinfo
  {author} {\bibfnamefont {P.}~\bibnamefont {Hansmann}},\ }\bibfield  {title}
  {\bibinfo {title} {{Interorbital singlet pairing in
  ${\mathrm{Sr}}_{2}{\mathrm{RuO}}_{4}$: A Hund's superconductor}},\ }\href
  {https://doi.org/10.1103/PhysRevB.105.155101} {\bibfield  {journal} {\bibinfo
   {journal} {Phys. Rev. B}\ }\textbf {\bibinfo {volume} {105}},\ \bibinfo
  {pages} {155101} (\bibinfo {year} {2022})}\BibitemShut {NoStop}%
\bibitem [{\citenamefont {Shorikov}\ \emph {et~al.}(2022)\citenamefont
  {Shorikov}, \citenamefont {Novoselov}, \citenamefont {Korotin},\ and\
  \citenamefont {Anisimov}}]{Shorikov2022}%
  \BibitemOpen
  \bibfield  {author} {\bibinfo {author} {\bibfnamefont {A.~O.}\ \bibnamefont
  {Shorikov}}, \bibinfo {author} {\bibfnamefont {D.~Y.}\ \bibnamefont
  {Novoselov}}, \bibinfo {author} {\bibfnamefont {D.~M.}\ \bibnamefont
  {Korotin}},\ and\ \bibinfo {author} {\bibfnamefont {V.~I.}\ \bibnamefont
  {Anisimov}},\ }\bibfield  {title} {\bibinfo {title} {{Orbital Selective
  Localization Enhancement in Ca$_{2 – x}$Sr$_x$RuO$_4$}},\ }\href
  {https://doi.org/10.1134/s0021364022602391} {\bibfield  {journal} {\bibinfo
  {journal} {JETP Letters}\ }\textbf {\bibinfo {volume} {116}},\ \bibinfo
  {pages} {798–803} (\bibinfo {year} {2022})}\BibitemShut {NoStop}%
\bibitem [{\citenamefont {Suzuki}\ \emph {et~al.}(2023)\citenamefont {Suzuki},
  \citenamefont {Wang}, \citenamefont {Bertinshaw}, \citenamefont {Strand},
  \citenamefont {K\"{a}ser}, \citenamefont {Krautloher}, \citenamefont {Yang},
  \citenamefont {Wentzell}, \citenamefont {Parcollet}, \citenamefont
  {Jerzembeck}, \citenamefont {Kikugawa}, \citenamefont {Mackenzie},
  \citenamefont {Georges}, \citenamefont {Hansmann}, \citenamefont
  {Gretarsson},\ and\ \citenamefont {Keimer}}]{Suzuki2023}%
  \BibitemOpen
  \bibfield  {author} {\bibinfo {author} {\bibfnamefont {H.}~\bibnamefont
  {Suzuki}}, \bibinfo {author} {\bibfnamefont {L.}~\bibnamefont {Wang}},
  \bibinfo {author} {\bibfnamefont {J.}~\bibnamefont {Bertinshaw}}, \bibinfo
  {author} {\bibfnamefont {H.~U.~R.}\ \bibnamefont {Strand}}, \bibinfo {author}
  {\bibfnamefont {S.}~\bibnamefont {K\"{a}ser}}, \bibinfo {author}
  {\bibfnamefont {M.}~\bibnamefont {Krautloher}}, \bibinfo {author}
  {\bibfnamefont {Z.}~\bibnamefont {Yang}}, \bibinfo {author} {\bibfnamefont
  {N.}~\bibnamefont {Wentzell}}, \bibinfo {author} {\bibfnamefont
  {O.}~\bibnamefont {Parcollet}}, \bibinfo {author} {\bibfnamefont
  {F.}~\bibnamefont {Jerzembeck}}, \bibinfo {author} {\bibfnamefont
  {N.}~\bibnamefont {Kikugawa}}, \bibinfo {author} {\bibfnamefont {A.~P.}\
  \bibnamefont {Mackenzie}}, \bibinfo {author} {\bibfnamefont {A.}~\bibnamefont
  {Georges}}, \bibinfo {author} {\bibfnamefont {P.}~\bibnamefont {Hansmann}},
  \bibinfo {author} {\bibfnamefont {H.}~\bibnamefont {Gretarsson}},\ and\
  \bibinfo {author} {\bibfnamefont {B.}~\bibnamefont {Keimer}},\ }\bibfield
  {title} {\bibinfo {title} {{Distinct spin and orbital dynamics in
  Sr$_2$RuO$_4$}},\ }\bibfield  {journal} {\bibinfo  {journal} {Nature
  Communications}\ }\textbf {\bibinfo {volume} {14}},\ \href
  {https://doi.org/10.1038/s41467-023-42804-3} {10.1038/s41467-023-42804-3}
  (\bibinfo {year} {2023})\BibitemShut {NoStop}%
\bibitem [{\citenamefont {Moon}(2023)}]{moon23}%
  \BibitemOpen
  \bibfield  {author} {\bibinfo {author} {\bibfnamefont {C.-Y.}\ \bibnamefont
  {Moon}},\ }\bibfield  {title} {\bibinfo {title} {{Effects of orbital
  selective dynamical correlation on the spin susceptibility and
  superconducting symmetries in ${\mathrm{Sr}}_{2}{\mathrm{RuO}}_{4}$}},\
  }\href {https://doi.org/10.1103/PhysRevResearch.5.L022058} {\bibfield
  {journal} {\bibinfo  {journal} {Phys. Rev. Res.}\ }\textbf {\bibinfo {volume}
  {5}},\ \bibinfo {pages} {L022058} (\bibinfo {year} {2023})}\BibitemShut
  {NoStop}%
\bibitem [{\citenamefont {Blesio}\ \emph {et~al.}(2024)\citenamefont {Blesio},
  \citenamefont {Beck}, \citenamefont {Gingras}, \citenamefont {Georges},\ and\
  \citenamefont {Mravlje}}]{blesio24}%
  \BibitemOpen
  \bibfield  {author} {\bibinfo {author} {\bibfnamefont {G.}~\bibnamefont
  {Blesio}}, \bibinfo {author} {\bibfnamefont {S.}~\bibnamefont {Beck}},
  \bibinfo {author} {\bibfnamefont {O.}~\bibnamefont {Gingras}}, \bibinfo
  {author} {\bibfnamefont {A.}~\bibnamefont {Georges}},\ and\ \bibinfo {author}
  {\bibfnamefont {J.}~\bibnamefont {Mravlje}},\ }\bibfield  {title} {\bibinfo
  {title} {{Signatures of Hund metal and finite-frequency nesting in
  ${\mathrm{Sr}}_{2}{\mathrm{RuO}}_{4}$ revealed by electronic Raman
  scattering}},\ }\href {https://doi.org/10.1103/PhysRevResearch.6.023124}
  {\bibfield  {journal} {\bibinfo  {journal} {Phys. Rev. Res.}\ }\textbf
  {\bibinfo {volume} {6}},\ \bibinfo {pages} {023124} (\bibinfo {year}
  {2024})}\BibitemShut {NoStop}%
\bibitem [{\citenamefont {Deng}\ \emph {et~al.}(2016)\citenamefont {Deng},
  \citenamefont {Haule},\ and\ \citenamefont
  {Kotliar}}]{kotliar_transport_in_ruthenates}%
  \BibitemOpen
  \bibfield  {author} {\bibinfo {author} {\bibfnamefont {X.}~\bibnamefont
  {Deng}}, \bibinfo {author} {\bibfnamefont {K.}~\bibnamefont {Haule}},\ and\
  \bibinfo {author} {\bibfnamefont {G.}~\bibnamefont {Kotliar}},\ }\bibfield
  {title} {\bibinfo {title} {{Transport Properties of Metallic Ruthenates: A
  $\mathrm{DFT}+\mathrm{DMFT}$ Investigation}},\ }\href@noop {} {\bibfield
  {journal} {\bibinfo  {journal} {Phys. Rev. Lett.}\ }\textbf {\bibinfo
  {volume} {116}},\ \bibinfo {pages} {256401} (\bibinfo {year}
  {2016})}\BibitemShut {NoStop}%
\bibitem [{\citenamefont {Werner}\ and\ \citenamefont
  {Millis}(2007)}]{werner07}%
  \BibitemOpen
  \bibfield  {author} {\bibinfo {author} {\bibfnamefont {P.}~\bibnamefont
  {Werner}}\ and\ \bibinfo {author} {\bibfnamefont {A.~J.}\ \bibnamefont
  {Millis}},\ }\bibfield  {title} {\bibinfo {title} {{High-Spin to Low-Spin and
  Orbital Polarization Transitions in Multiorbital Mott Systems}},\ }\href
  {https://doi.org/10.1103/PhysRevLett.99.126405} {\bibfield  {journal}
  {\bibinfo  {journal} {Phys. Rev. Lett.}\ }\textbf {\bibinfo {volume} {99}},\
  \bibinfo {pages} {126405} (\bibinfo {year} {2007})}\BibitemShut {NoStop}%
\bibitem [{\citenamefont {Kune\ifmmode~\check{s}\else \v{s}\fi{}}\ and\
  \citenamefont {Augustinsk\'y}(2014)}]{kunes_augustinsky_2014}%
  \BibitemOpen
  \bibfield  {author} {\bibinfo {author} {\bibfnamefont {J.}~\bibnamefont
  {Kune\ifmmode~\check{s}\else \v{s}\fi{}}}\ and\ \bibinfo {author}
  {\bibfnamefont {P.}~\bibnamefont {Augustinsk\'y}},\ }\bibfield  {title}
  {\bibinfo {title} {{Excitonic instability at the spin-state transition in the
  two-band Hubbard model}},\ }\href
  {https://doi.org/10.1103/PhysRevB.89.115134} {\bibfield  {journal} {\bibinfo
  {journal} {Phys. Rev. B}\ }\textbf {\bibinfo {volume} {89}},\ \bibinfo
  {pages} {115134} (\bibinfo {year} {2014})}\BibitemShut {NoStop}%
\bibitem [{\citenamefont {Kune\ifmmode~\check{s}\else
  \v{s}\fi{}}(2014)}]{kunes2014}%
  \BibitemOpen
  \bibfield  {author} {\bibinfo {author} {\bibfnamefont {J.}~\bibnamefont
  {Kune\ifmmode~\check{s}\else \v{s}\fi{}}},\ }\bibfield  {title} {\bibinfo
  {title} {{Phase diagram of exciton condensate in doped two-band Hubbard
  model}},\ }\href {https://doi.org/10.1103/PhysRevB.90.235140} {\bibfield
  {journal} {\bibinfo  {journal} {Phys. Rev. B}\ }\textbf {\bibinfo {volume}
  {90}},\ \bibinfo {pages} {235140} (\bibinfo {year} {2014})}\BibitemShut
  {NoStop}%
\bibitem [{\citenamefont {Hoshino}\ and\ \citenamefont
  {Werner}(2016)}]{hoshino16}%
  \BibitemOpen
  \bibfield  {author} {\bibinfo {author} {\bibfnamefont {S.}~\bibnamefont
  {Hoshino}}\ and\ \bibinfo {author} {\bibfnamefont {P.}~\bibnamefont
  {Werner}},\ }\bibfield  {title} {\bibinfo {title} {{Electronic orders in
  multiorbital Hubbard models with lifted orbital degeneracy}},\ }\href
  {https://doi.org/10.1103/PhysRevB.93.155161} {\bibfield  {journal} {\bibinfo
  {journal} {Phys. Rev. B}\ }\textbf {\bibinfo {volume} {93}},\ \bibinfo
  {pages} {155161} (\bibinfo {year} {2016})}\BibitemShut {NoStop}%
\bibitem [{\citenamefont {Georges}\ \emph {et~al.}(2013)\citenamefont
  {Georges}, \citenamefont {de{\textquotesingle} Medici},\ and\ \citenamefont
  {Mravlje}}]{hund_coupling_jernej}%
  \BibitemOpen
  \bibfield  {author} {\bibinfo {author} {\bibfnamefont {A.}~\bibnamefont
  {Georges}}, \bibinfo {author} {\bibfnamefont {L.}~\bibnamefont
  {de{\textquotesingle} Medici}},\ and\ \bibinfo {author} {\bibfnamefont
  {J.}~\bibnamefont {Mravlje}},\ }\bibfield  {title} {\bibinfo {title} {{Strong
  Correlations from Hund's Coupling}},\ }\href@noop {} {\bibfield  {journal}
  {\bibinfo  {journal} {Annual Review of Condensed Matter Physics}\ }\textbf
  {\bibinfo {volume} {4}},\ \bibinfo {pages} {137} (\bibinfo {year}
  {2013})}\BibitemShut {NoStop}%
\bibitem [{\citenamefont {Gull}\ \emph {et~al.}(2011)\citenamefont {Gull},
  \citenamefont {Millis}, \citenamefont {Lichtenstein}, \citenamefont
  {Rubtsov}, \citenamefont {Troyer},\ and\ \citenamefont
  {Werner}}]{cthyb_long}%
  \BibitemOpen
  \bibfield  {author} {\bibinfo {author} {\bibfnamefont {E.}~\bibnamefont
  {Gull}}, \bibinfo {author} {\bibfnamefont {A.~J.}\ \bibnamefont {Millis}},
  \bibinfo {author} {\bibfnamefont {A.~I.}\ \bibnamefont {Lichtenstein}},
  \bibinfo {author} {\bibfnamefont {A.~N.}\ \bibnamefont {Rubtsov}}, \bibinfo
  {author} {\bibfnamefont {M.}~\bibnamefont {Troyer}},\ and\ \bibinfo {author}
  {\bibfnamefont {P.}~\bibnamefont {Werner}},\ }\bibfield  {title} {\bibinfo
  {title} {{Continuous-time Monte Carlo methods for quantum impurity models}},\
  }\href@noop {} {\bibfield  {journal} {\bibinfo  {journal} {Reviews of Modern
  Physics}\ }\textbf {\bibinfo {volume} {83}},\ \bibinfo {pages} {349}
  (\bibinfo {year} {2011})}\BibitemShut {NoStop}%
\bibitem [{\citenamefont {Seth}\ \emph {et~al.}(2016)\citenamefont {Seth},
  \citenamefont {Krivenko}, \citenamefont {Ferrero},\ and\ \citenamefont
  {Parcollet}}]{cthyb_triqs}%
  \BibitemOpen
  \bibfield  {author} {\bibinfo {author} {\bibfnamefont {P.}~\bibnamefont
  {Seth}}, \bibinfo {author} {\bibfnamefont {I.}~\bibnamefont {Krivenko}},
  \bibinfo {author} {\bibfnamefont {M.}~\bibnamefont {Ferrero}},\ and\ \bibinfo
  {author} {\bibfnamefont {O.}~\bibnamefont {Parcollet}},\ }\bibfield  {title}
  {\bibinfo {title} {{TRIQS/CTHYB: A continuous-time quantum Monte Carlo
  hybridisation expansion solver for quantum impurity problems}},\ }\href@noop
  {} {\bibfield  {journal} {\bibinfo  {journal} {Computer Physics
  Communications}\ }\textbf {\bibinfo {volume} {200}},\ \bibinfo {pages} {274 }
  (\bibinfo {year} {2016})}\BibitemShut {NoStop}%
\bibitem [{\citenamefont {Parcollet}\ \emph {et~al.}(2015)\citenamefont
  {Parcollet}, \citenamefont {Ferrero}, \citenamefont {Ayral}, \citenamefont
  {Hafermann}, \citenamefont {Krivenko}, \citenamefont {Messio},\ and\
  \citenamefont {Seth}}]{triqs}%
  \BibitemOpen
  \bibfield  {author} {\bibinfo {author} {\bibfnamefont {O.}~\bibnamefont
  {Parcollet}}, \bibinfo {author} {\bibfnamefont {M.}~\bibnamefont {Ferrero}},
  \bibinfo {author} {\bibfnamefont {T.}~\bibnamefont {Ayral}}, \bibinfo
  {author} {\bibfnamefont {H.}~\bibnamefont {Hafermann}}, \bibinfo {author}
  {\bibfnamefont {I.}~\bibnamefont {Krivenko}}, \bibinfo {author}
  {\bibfnamefont {L.}~\bibnamefont {Messio}},\ and\ \bibinfo {author}
  {\bibfnamefont {P.}~\bibnamefont {Seth}},\ }\bibfield  {title} {\bibinfo
  {title} {{TRIQS: A toolbox for research on interacting quantum systems}},\
  }\href@noop {} {\bibfield  {journal} {\bibinfo  {journal} {Computer Physics
  Communications}\ }\textbf {\bibinfo {volume} {196}},\ \bibinfo {pages} {398 }
  (\bibinfo {year} {2015})}\BibitemShut {NoStop}%
\bibitem [{\citenamefont {Beach}(2004)}]{stochME}%
  \BibitemOpen
  \bibfield  {author} {\bibinfo {author} {\bibfnamefont {K.~S.~D.}\
  \bibnamefont {Beach}},\ }\href@noop {} {\bibinfo {title} {{Identifying the
  maximum entropy method as a special limit of stochastic analytic
  continuation}}} (\bibinfo {year} {2004}),\ \Eprint
  {https://arxiv.org/abs/cond-mat/0403055} {arXiv:cond-mat/0403055
  [cond-mat.str-el]} \BibitemShut {NoStop}%
\bibitem [{\citenamefont {Arsenault}\ and\ \citenamefont
  {Tremblay}(2013)}]{arsenault}%
  \BibitemOpen
  \bibfield  {author} {\bibinfo {author} {\bibfnamefont {L.-F.}\ \bibnamefont
  {Arsenault}}\ and\ \bibinfo {author} {\bibfnamefont {A.-M.~S.}\ \bibnamefont
  {Tremblay}},\ }\bibfield  {title} {\bibinfo {title} {{Transport functions for
  hypercubic and Bethe lattices}},\ }\href@noop {} {\bibfield  {journal}
  {\bibinfo  {journal} {Physical Review B}\ }\textbf {\bibinfo {volume} {88}}
  (\bibinfo {year} {2013})}\BibitemShut {NoStop}%
\bibitem [{\citenamefont {Pruschke}\ \emph {et~al.}(1993)\citenamefont
  {Pruschke}, \citenamefont {Cox},\ and\ \citenamefont {Jarrell}}]{cox}%
  \BibitemOpen
  \bibfield  {author} {\bibinfo {author} {\bibfnamefont {T.}~\bibnamefont
  {Pruschke}}, \bibinfo {author} {\bibfnamefont {D.~L.}\ \bibnamefont {Cox}},\
  and\ \bibinfo {author} {\bibfnamefont {M.}~\bibnamefont {Jarrell}},\
  }\bibfield  {title} {\bibinfo {title} {{Hubbard model at infinite dimensions:
  Thermodynamic and transport properties}},\ }\href@noop {} {\bibfield
  {journal} {\bibinfo  {journal} {Physical Review B}\ }\textbf {\bibinfo
  {volume} {47}},\ \bibinfo {pages} {3553} (\bibinfo {year}
  {1993})}\BibitemShut {NoStop}%
\bibitem [{\citenamefont {Khurana}(1990)}]{vertexcorrections}%
  \BibitemOpen
  \bibfield  {author} {\bibinfo {author} {\bibfnamefont {A.}~\bibnamefont
  {Khurana}},\ }\bibfield  {title} {\bibinfo {title} {{Electrical conductivity
  in the infinite-dimensional Hubbard model}},\ }\href@noop {} {\bibfield
  {journal} {\bibinfo  {journal} {Phys. Rev. Lett.}\ }\textbf {\bibinfo
  {volume} {64}},\ \bibinfo {pages} {1990} (\bibinfo {year}
  {1990})}\BibitemShut {NoStop}%
\bibitem [{\citenamefont {Chung}\ and\ \citenamefont
  {Freericks}(1998)}]{chung}%
  \BibitemOpen
  \bibfield  {author} {\bibinfo {author} {\bibfnamefont {W.}~\bibnamefont
  {Chung}}\ and\ \bibinfo {author} {\bibfnamefont {J.~K.}\ \bibnamefont
  {Freericks}},\ }\bibfield  {title} {\bibinfo {title} {{Charge-transfer
  metal-insulator transitions in the spin-$\frac{1}{2}$ Falicov-Kimball
  model}},\ }\href {https://doi.org/10.1103/PhysRevB.57.11955} {\bibfield
  {journal} {\bibinfo  {journal} {Phys. Rev. B}\ }\textbf {\bibinfo {volume}
  {57}},\ \bibinfo {pages} {11955} (\bibinfo {year} {1998})}\BibitemShut
  {NoStop}%
\bibitem [{\citenamefont {Piessens}\ \emph {et~al.}(2012)\citenamefont
  {Piessens}, \citenamefont {de~Doncker-Kapenga}, \citenamefont
  {{\"U}berhuber},\ and\ \citenamefont {Kahaner}}]{scipy}%
  \BibitemOpen
  \bibfield  {author} {\bibinfo {author} {\bibfnamefont {R.}~\bibnamefont
  {Piessens}}, \bibinfo {author} {\bibfnamefont {E.}~\bibnamefont
  {de~Doncker-Kapenga}}, \bibinfo {author} {\bibfnamefont {C.~W.}\ \bibnamefont
  {{\"U}berhuber}},\ and\ \bibinfo {author} {\bibfnamefont {D.~K.}\
  \bibnamefont {Kahaner}},\ }\href@noop {} {\emph {\bibinfo {title} {{Quadpack:
  a subroutine package for automatic integration}}}},\ Vol.~\bibinfo {volume}
  {1}\ (\bibinfo  {publisher} {Springer Science \& Business Media},\ \bibinfo
  {year} {2012})\BibitemShut {NoStop}%
\bibitem [{\citenamefont {Van~Vleck}(1932)}]{vanvleck_original}%
  \BibitemOpen
  \bibfield  {author} {\bibinfo {author} {\bibfnamefont {J.~H.}\ \bibnamefont
  {Van~Vleck}},\ }\href@noop {} {\emph {\bibinfo {title} {{The theory of
  electric and magnetic susceptibilities}}}}\ (\bibinfo  {publisher} {Oxford
  University Press},\ \bibinfo {year} {1932})\BibitemShut {NoStop}%
\bibitem [{\citenamefont {Mulak}(1986)}]{mulak}%
  \BibitemOpen
  \bibfield  {author} {\bibinfo {author} {\bibfnamefont {J.}~\bibnamefont
  {Mulak}},\ }\bibfield  {title} {\bibinfo {title} {{Crystal field effect and
  temperature dependence of paramagnetic susceptibility}},\ }\href@noop {}
  {\bibfield  {journal} {\bibinfo  {journal} {Journal of the Less Common
  Metals}\ }\textbf {\bibinfo {volume} {121}},\ \bibinfo {pages} {141}
  (\bibinfo {year} {1986})}\BibitemShut {NoStop}%
\bibitem [{\citenamefont {Koenig}\ and\ \citenamefont
  {Madeja}(1967)}]{vanvleck_koenig}%
  \BibitemOpen
  \bibfield  {author} {\bibinfo {author} {\bibfnamefont {E.}~\bibnamefont
  {Koenig}}\ and\ \bibinfo {author} {\bibfnamefont {K.}~\bibnamefont
  {Madeja}},\ }\bibfield  {title} {\bibinfo {title} {{5T2-1A1 Equilibriums in
  some iron(II)-bis(1,10-phenanthroline) complexes}},\ }\href@noop {}
  {\bibfield  {journal} {\bibinfo  {journal} {Inorganic Chemistry}\ }\textbf
  {\bibinfo {volume} {6}},\ \bibinfo {pages} {48} (\bibinfo {year}
  {1967})}\BibitemShut {NoStop}%
\bibitem [{\citenamefont {Deng}\ \emph {et~al.}(2012)\citenamefont {Deng},
  \citenamefont {Ferrero}, \citenamefont {Mravlje}, \citenamefont {Aichhorn},\
  and\ \citenamefont {Georges}}]{deng2012}%
  \BibitemOpen
  \bibfield  {author} {\bibinfo {author} {\bibfnamefont {X.}~\bibnamefont
  {Deng}}, \bibinfo {author} {\bibfnamefont {M.}~\bibnamefont {Ferrero}},
  \bibinfo {author} {\bibfnamefont {J.}~\bibnamefont {Mravlje}}, \bibinfo
  {author} {\bibfnamefont {M.}~\bibnamefont {Aichhorn}},\ and\ \bibinfo
  {author} {\bibfnamefont {A.}~\bibnamefont {Georges}},\ }\bibfield  {title}
  {\bibinfo {title} {{Hallmark of strong electronic correlations in
  LaNiO${}_{3}$: Photoemission kink and broadening of fully occupied bands}},\
  }\href@noop {} {\bibfield  {journal} {\bibinfo  {journal} {Phys. Rev. B}\
  }\textbf {\bibinfo {volume} {85}},\ \bibinfo {pages} {125137} (\bibinfo
  {year} {2012})}\BibitemShut {NoStop}%
\bibitem [{\citenamefont {Maeno}\ \emph {et~al.}(1997)\citenamefont {Maeno},
  \citenamefont {Yoshida}, \citenamefont {Hashimoto}, \citenamefont
  {Nishizaki}, \citenamefont {Ikeda}, \citenamefont {Nohara}, \citenamefont
  {Fujita}, \citenamefont {Mackenzie}, \citenamefont {Hussey}, \citenamefont
  {Bednorz},\ and\ \citenamefont {Lichtenberg}}]{Maeno1997}%
  \BibitemOpen
  \bibfield  {author} {\bibinfo {author} {\bibfnamefont {Y.}~\bibnamefont
  {Maeno}}, \bibinfo {author} {\bibfnamefont {K.}~\bibnamefont {Yoshida}},
  \bibinfo {author} {\bibfnamefont {H.}~\bibnamefont {Hashimoto}}, \bibinfo
  {author} {\bibfnamefont {S.}~\bibnamefont {Nishizaki}}, \bibinfo {author}
  {\bibfnamefont {S.-i.}\ \bibnamefont {Ikeda}}, \bibinfo {author}
  {\bibfnamefont {M.}~\bibnamefont {Nohara}}, \bibinfo {author} {\bibfnamefont
  {T.}~\bibnamefont {Fujita}}, \bibinfo {author} {\bibfnamefont {A.~P.}\
  \bibnamefont {Mackenzie}}, \bibinfo {author} {\bibfnamefont {N.~E.}\
  \bibnamefont {Hussey}}, \bibinfo {author} {\bibfnamefont {J.~G.}\
  \bibnamefont {Bednorz}},\ and\ \bibinfo {author} {\bibfnamefont
  {F.}~\bibnamefont {Lichtenberg}},\ }\bibfield  {title} {\bibinfo {title}
  {{Two-Dimensional Fermi Liquid Behavior of the Superconductor
  Sr$_2$RuO$_4$}},\ }\href {https://doi.org/10.1143/jpsj.66.1405} {\bibfield
  {journal} {\bibinfo  {journal} {Journal of the Physical Society of Japan}\
  }\textbf {\bibinfo {volume} {66}},\ \bibinfo {pages} {1405–1408} (\bibinfo
  {year} {1997})}\BibitemShut {NoStop}%
\bibitem [{dat()}]{datarep}%
  \BibitemOpen
  \href@noop {} {}\bibinfo {note} {Graz University of Technology data
  repository, https://doi.org/10.3217/v82n3-nhk43}\BibitemShut {NoStop}%
\end{thebibliography}%

\end{document}